\documentclass[conference]{IEEEtran}
\ifCLASSINFOpdf
\else
\fi
\usepackage{algorithm}
\usepackage{algpseudocode}
\usepackage{hyperref}
\usepackage{caption}
\usepackage{subcaption}
\usepackage{amsmath}
\usepackage{graphicx}
\usepackage{soul}
\usepackage{xcolor}
\usepackage[normalem]{ulem}

\newcommand{\proj}{\textsc{Repttack}}

\begin{document}
% \title{When Users Have Control over Scheduler Outputs: Weak-Points of Cloud Schedulers Leading to Risk of Co-Location Attacks}
\title{{\proj}: Exploiting Cloud Schedulers to Guide Co-Location Attacks}

% \author{
% \IEEEauthorblockN{Chongzhou Fang}
% \IEEEauthorblockA{University of California, Davis\\
% czfang@ucdavis.edu}
% \and
% \IEEEauthorblockN{Han Wang}
% \IEEEauthorblockA{University of California, Davis\\
% hjlwang@ucdavis.edu}
% \and
% \IEEEauthorblockN{Najmeh Nazari}
% \IEEEauthorblockA{University of California, Davis\\
% nnazaribavarsad@ucdavis.edu}
% \and
% \IEEEauthorblockN{Behnam Omidi}
% \IEEEauthorblockA{George Mason University\\
% bomidi@masonlive.gmu.edu}
% \IEEEauthorblockN{Avesta Sasan}
% \IEEEauthorblockA{George Mason University\\
% asasan@gmu.edu}
% \and
% \IEEEauthorblockN{Houman Homayoun}
% \IEEEauthorblockA{University of California, Davis\\
% hhomayoun@ucdavis.edu}
% }

\author{\IEEEauthorblockN{Chongzhou Fang\IEEEauthorrefmark{1},
Han Wang\IEEEauthorrefmark{1},
Najmeh Nazari\IEEEauthorrefmark{1}, 
Behnam Omidi\IEEEauthorrefmark{2},
Avesta Sasan\IEEEauthorrefmark{1},\\
Khaled N. Khasawneh\IEEEauthorrefmark{2},
Setareh Rafatirad\IEEEauthorrefmark{1},and 
Houman Homayoun\IEEEauthorrefmark{1}}
\IEEEauthorblockA{\IEEEauthorrefmark{1}
University of California, Davis\\
Email: \{czfang,hjlwang,nnazaribavarsad,asasan,srafatirad,hhomayoun\}@ucdavis.edu}
\IEEEauthorblockA{\IEEEauthorrefmark{2}
George Mason University\\
Email: \{bomidi,kkhasawn\}@gmu.edu}
}


%Telephone: (800) 555--1212, Fax: (888) 555--1212}


\IEEEoverridecommandlockouts
\makeatletter\def\@IEEEpubidpullup{6.5\baselineskip}\makeatother
\IEEEpubid{\parbox{\columnwidth}{
    Network and Distributed Systems Security (NDSS) Symposium 2022\\
    27 February - 3 March 2022\\
    ISBN 1-891562-66-5\\
    https://dx.doi.org/10.14722/ndss.2022.23149\\
    www.ndss-symposium.org
}
\hspace{\columnsep}\makebox[\columnwidth]{}}

\maketitle

\begin{abstract}
%\jane{Not sure the focus will be datacenters or cloud?}
Cloud computing paradigms have emerged as a major facility to store and process the massive data produced by various business units, public organizations, Internet-of-Things (IoT), and cyber-physical systems (CPS). To meet users' performance requirements while maximizing resource utilization to achieve cost-efficiency, cloud administrators leverage schedulers to orchestrate tasks to different physical nodes and allow applications from different users to share the same physical node. On the other hand, micro-architectural attacks, e.g, side-channel attacks, transient execution attacks, and Rowhammer attacks, exploit the shared resources to compromise the confidentiality/integrity of a co-located victim application. Since co-location is an essential requirement for micro-architectural attacks, in this work, we investigate whether attackers can exploit the cloud schedulers to satisfy the co-location requirement of the micro-architectural attacks. Specifically, in this paper, we comprehensively analyze if attackers can influence the scheduling process of cloud schedulers to co-locate with specific targeted applications in the cloud. Our analysis shows that for cloud schedulers that allow users to submit application requirements, an attacker can carefully select the attacker's application requirements to influence the scheduler to co-locate it with a targeted victim application. We call such attack \textit{Rep}lication A\textit{ttack} ({\proj}). Our experimental results, in both a simulated cluster environment and a real cluster, show similar trends; a single attack instance can reach up to $50\%$ co-location rate (probability of co-location) and with only $5$ instances the co-location rate can reach up to $80\%$ in a heterogeneous cloud. Furthermore, we propose and evaluate a mitigation strategy that can help defend against {\proj}. We believe that our results highlight the fact that schedulers in multi-user clusters need to be more carefully designed with security in mind, and the process of making scheduling decisions should involve as little user-defined information as possible.
\end{abstract}



% \fi
%%\IEEEpeerreviewmaketitle

% \hl{intro needs}

\section{Introduction}
\label{secIntro}

In the last two decades, the computing paradigms have experienced a tremendous change, with vast amounts of data being amassed by various business units and public organizations. This rate of data amassing is further fueled by the advancements in the Internet-of-Things (IoT) and cyber-physical systems (CPS) equipped with networking capabilities and rapid movement towards paperless organizations~\cite{subramanian2018recent}. %An increasing number of organizations and companies build their datacenters, hoping to provide more powerful computing resources in an easy-to-maintain manner.
For security consideration, virtual machines (VMs)~\cite{raj2009resource} or containers~\cite{pahl2017cloud} are utilized to achieve software level isolation between different users. VMs or containers are called instances in this work. To better manage the submitted workloads and improve utilization rate automatically, scheduling algorithms are proposed to help decide resources assigned to each task submitted and which physical node a task should reside. A number of schedulers, including Apollo~\cite{boutin2014apollo}, YARN~\cite{vavilapalli2013apache} achieve great success in the industry. However, the current design of the scheduling algorithms focuses on enhancing performance, utilization rate, and load-balancing without the security consideration, which may bring new vulnerability as we explore in this work.

Micro-architectural attacks, such as side-channel attacks~\cite{yarom2016mastik}, covert-channel attacks~\cite{naghibijouybari2017constructing}, Rowhammer~\cite{kim2014flipping}, and transient execution attacks~\cite{kocher2019spectre,lipp2018meltdown,koruyeh2018spectre}, exploit shared resources to compromise the confidentiality/integrity of a co-located victim application/instances. Thus, attackers usually target clouds to launch micro-architectural attacks since clouds allow attackers' instances to share resources with other users' instances to achieve cost-efficiency. However, a major challenge in launching micro-architectural attacks in clouds is the ability of the attack instance to co-locate with a targeted victim. While prior work showed that co-located applications can be identified using network probing~\cite{ristenpart2009hey} or side-channel analysis~\cite{zhang2011homealone,inci2016co}, the only way to increase the co-location probability of the attack instance with a targeted victim is by launching a large number of attack instances (possibly with the assistance of placement timing locality), i.e., brute-force approach, and hope that the attack instance co-locates with a targeted victim. Nonetheless, to defend against the brute-force, cloud managers can limit the number of instances a user can issue within a time frame.

%Since the applications from different users can be placed on the same physical node, this resource sharing opens up chances for various types of attacks, such as side-channel attacks~\cite{yarom2016mastik}, Rowhammer~\cite{gruss2016rowhammer}, etc. According to~\cite{ristenpart2009hey,inci2015seriously}, an attacker can first manage to co-locate his malicious instances on the same physical machine where the victim application is running on. Then the attacker identifies whether the co-location succeeds or not by launching network probing~\cite{ristenpart2009hey} or side-channel analysis~\cite{zhang2011homealone,inci2016co}. Once the victim is identified running on the same machine with attackers, malicious activities are initiated, which can be side-channel attacks~\cite{disselkoen2017prime+}, Rowhammer~\cite{gruss2016rowhammer}, etc. Such attacks enable the compromising of clouds in terms of availability~\cite{gupta2013vm}, information privacy~\cite{}, confidentiality~\cite{}. Since a wide range of commercial processors share similar vulnerabilities~\cite{lyu2018survey}, the risks commonly exist on cloud servers. However, cloud managers can detect and mitigate the attacks by measuring and limiting the number of instances a user can issue within a time frame.

In this work, we explore whether attackers can exploit the cloud schedulers to force co-locating an attack instance with a targeted victim. In particular, we target schedulers that allow users to provide specifications of the submitted instances. These specifications influence the placement decisions of the submitted instances. Our attack strategy is to camouflage the victim specifications for the attack instance to achieve co-location. Thus, allowing the attacker to achieve co-location by launching a few instances, which can't be defeated by limiting the number of instances a user can submit.
%is to replicate such information, disguise as a normal user and submit the attack instance to achieve co-location. Due to the stealthy characteristics, it is not trivial for cloud administrators to detect it by monitoring instance requests.

To achieve this, we analyze the parameters that schedulers take as input to decide the location of submitted workloads and explore the possibility of manipulating the scheduling outcome for malicious instances by controlling the parameters. In particular, this work investigates the effects of certain parameters by analyzing the co-location rate of malicious instances and victim instances under different parameter settings, and validate that they contribute to vulnerabilities of schedulers. 
%In particular, this work investigates the input parameters extensively by analyzing the co-location rate of malicious instance and victim instance under different parameter settings. Then, the parameters contributing most to the final co-location decision are selected and identified as "vulnerable parameters". 
We choose open-source scheduling strategies based on the filter-score scheduling method as examples and find that such scheduling algorithms leverage a special feature called affinity that allows users to fine-tune the scheduling output makes the scheduler more vulnerable. We find that the feature exists in schedulers of cloud architecture like Kubernates~\cite{kubernetes}, Openstack~\cite{openstack}, OpenNebula~\cite{opennebula}, etc. and are used in a recent IoT system~\cite{santos2019towards}. Hence we believe the risk we identify in this paper applies to real-world scenarios. Therefore, showing that security needs to be considered when designing cloud schedulers on par with performance and cost-efficacy.

%To further demonstrate the threat imposed by the venerability in scheduling algorithms, we launch Flush+Reload on deep learning application, Prime+Probe on encryption application, DDoS on XX respectively and evaluate the success rate. We find that XX.

Lastly, based on our findings, we provide a strategy to mitigate the vulnerability by making a small change to the scheduler and show how bringing randomness in the scheduling process can help avoid the security risks. 
%several scheduling design to mitigate the venerability and  non-malicious users and scheduler designers regarding how to utilize or how to avoid the security risks. 
% 

Our contributions in this paper are:
\begin{itemize}%[$$]
\item We identify the security risks of manipulating schedulers to achieve co-location with victim instances with less than $5$ malicious instances requested.
% \item We propose a framework to identify
\item Experiments that simulate the process of deploying applications in real-world scenarios have been done in both simulation and CloudLab cluster~\cite{duplyakin2019design} to identify the vulnerable features.
\item We find that the use of affinity features in filter-score schedulers can induce security problems and propose {\proj}, which is an attack method that utilizes such security vulnerabilities. We show that with these features used, {\proj} can enable malicious users to achieve a relatively high co-location rate in a heterogeneous cluster.
\item Based on our findings, we conclude several guidelines for attackers, non-malicious users, and scheduler designers regarding how to utilize or how to avoid the security risks.
\item We propose a mitigation method and validate it in our simulation. It shows how bringing randomness to the scheduling process can help to defend against co-location attacks.
%We also provide other interesting findings in this paper.
 
\end{itemize}

Our work focuses on achieving co-location to satisfy the prerequisite step of launching micro-architectural attacks and considers the implementation of micro-architectural attacks on a specific hardware as an orthogonal topic to this work. Attackers can utilize micro-architectural attack methods reported in literature (e.g. \cite{disselkoen2017prime+,gruss2016flush+,kocher2019spectre,lipp2018meltdown}) to craft their own attack instances.

The remainder of this paper is organized as follows. In Section \ref{SecBackground}, we will provide a brief introduction to cloud schedulers and some related background knowledge on micro-architectural attacks. In Section \ref{SecThreat}, we provide the assumptions we make in this study. Section \ref{SecMethod} summarizes our attack strategy, and Section \ref{SecExperiments} presents corresponding results and analysis. Section \ref{SecMitigation} describes our mitigation strategy and presents related data. In Section \ref{SecDiscussion} we provide a discussion that focuses more on the big picture of this field and points out our future direction. Related literature will be reviewed in Section \ref{SecRelatedWk}, and we conclude the entire paper in Section \ref{SecConclusion}.

\section{Background}\label{SecBackground}
This section first introduces cloud schedulers followed by an overview of micro-architectural attacks.

\subsection{Cloud Schedulers}
A scheduler is an essential component of a distributed computing system. It orchestrates resources, assigns resources in the system, and makes placement decisions that satisfy user requirements with fairness for all users~\cite{schwarzkopf2013omega}. In this paper, we consider the scheduler as the component that decides how a user instance should be placed and provide resources. The scheduling algorithm receives user requirements on placement and assigns user instances to physical nodes that can satisfy users' needs. We summarize the working procedure of cloud schedulers as: First, users submit needed scheduling constraints (needed resources, requirements on nodes, etc.) to the cloud scheduler. Then, the cloud scheduler considers all available resources in the cluster as well as user constraints and assigns a node with enough resources that match the required constraints to that user's instance. Different user instances can be allocated to the same node and share the available hardware resources. A more detailed elaboration of the scheduler we target is provided in Section \ref{SecMethodTarget}.

Cloud schedulers are extensively studied in academia and widely used in industry~\cite{vavilapalli2013apache,schwarzkopf2013omega,boutin2014apollo,karanasos2015mercury,verma2015large,hadary2020protean}. Open-source cloud architectures, including schedulers as components, are also utilized in the industry, e.g., OpenStack \cite{openstack} is deployed in multiple industrial companies (ELASTX, China Mobile \cite{openstack}, for example). However, the previous focus of academia and industry is performance, load-balancing and cost. In contrast, we focus on scheduler security in this work. 

\subsection{Micro-architectural Attacks}
Micro-architectural attacks exploit the functionality, optimization, or the physical imperfections of shared resources to compromise the confidentiality/integrity of a victim application. Furthermore, micro-architectural attacks are software-based attacks, i.e., can be launched remotely, and can be categorized into the following four categories:  

\noindent
\textbf{(1) Side-channel attack:} is one emerging category of techniques in the field of computer security. Side-channel attack utilizes the data-dependent effects of computation/operations/optimizations, which are also referred to as side-channels, to obtain sensitive information, like secret keys, password and etc. from target software systems~\cite{yarom2014flush+,gruss2016flush+}. Researchers have shown that by utilizing side-channel information such as the access time to caches~\cite{yarom2014flush+,gruss2016flush+,disselkoen2017prime+} information can be leaked from co-located applications. Compared to traditional attack techniques, side-channel attack is more stealthy and harder to defend against since it targets hardware design weak points.

\noindent
\textbf{(2) Transient execution attacks:} exploits flaws in micro-architecture to execute operations that should not have happened~\cite{canella2020evolution}. For example, in 2018, Meltdown attack is proposed~\cite{lipp2018meltdown}, which utilizes out-of-order execution as a new side-channel and enables attackers to read arbitrary physical memory addresses of other processes. A more recent similar work, Spectre Attack~\cite{kocher2019spectre}, additionally utilizes branch prediction to mistrain branch predictors and lead to the execution of illegal paths. Spectre can be applied to a wider range of processors. These two attack technologies affect a wide range of processors~\cite{meltdownspectre} since they target speculative execution, which has been deployed in commercial processors for decades. Moreover, despite the advances in the proposed and deployed defenses against transient execution attacks~\cite{yan2018invisispec,xiong2020survey,islam2020nd,ARMSpecAnalysis_whitepaper,koruyeh2020speccfi,Turner2018retpoline,Intel2018retpoline,khasawneh2019safespec}, they are still limited~\cite{canella2020evolutiond}.

\noindent
\textbf{(3) Rowhammer attack:} Current DRAM chips are designed to be physically denser to increase capacity and reduce energy consumption. Though there are benefits of such a design, it leverages smaller cells that can only hold a lower charge, reducing the noise threshold. At the same time, the smaller distance between DRAM cells causes electromagnetic coupling effects, which means that a small amount of charge is leaked when an adjacent cell is read. Kim \textit{et. al}~\cite{kim2014flipping} presents that the frequent read of one DRAM cell can cause bit-flip in a neighbor row, which can be used to build a DRAM-based attack, termed as ``rowhammer".

% \noindent
% \textbf{(3) Rowhammer attack:} exploits the Rowhammer bug in DRAM memory access~\cite{kim2014flipping,seaborn2015exploiting}, and can cause bit flips and further obtain privileged control of the system. 

\noindent
\textbf{(4) Faults attack:} %plundervolt and clkscrew attacks.
employs vulnerabilities of frequency/voltage adjusting mechanisms existing in computer systems and induces faults in program execution. For example, CLKscrew~\cite{tang2017clkscrew} targets the vulnerabilities in power management in ARM devices. Plundervolt~\cite{murdock2020plundervolt} targets the voltage scaling feature in Intel processors. They both abuse such flaws to induce faults that can be further utilized to reveal security keys, etc.

Since servers in the cloud use similar architectures, the same vulnerabilities of micro-architectural attacks are also a threat to cloud providers and users~\cite{ristenpart2009hey}. But to initiate side-channel attacks in the cloud, attackers must first achieve co-location with victims. This makes deploying micro-architectural attacks in the cloud an interesting problem since it involves both seizing control of cloud schedulers and exploiting weak points in cloud infrastructure hardware. Our focus in this paper is on the former step. There are off-the-shelf implementation of these attacks hence it is relatively easy to launch micro-architectural attacks once co-location is achieved.% and it will be our future work to demonstrate how micro-architectural attacks can be stealthily deployed.

\section{Threat Model}\label{SecThreat}
In this study, attackers' goal is to exploit scheduler vulnerabilities and force the scheduler to locate at least one of the attack instances to the same physical node with the victim instance. As introduced in Section \ref{secIntro}, a number of micro-architectural attacks demand co-location to compromise users' privacy~\cite{gruss2016flush+} and system availability~\cite{jang2017sgx}. This work demonstrates that attackers are able to leverage cloud schedulers to co-locate with target victim applications. We target a feature in filter-score scheduler that is used in popular cloud architectures including Kubernetes~\cite{kubernetes}, OpenStack~\cite{openstack}, OpenNebula~\cite{opennebula}, etc.
%  This step is the prerequisite of micro-architectural attacks. After successfully co-locating with the victim instance, micro-architectural attacks can be launched. In this paper, we only focus on the co-location step.

Furthermore, we consider cloud providers as trusted service providers and have no additional information about their users; hence, they will consider all users as equivalent and not differentiate users according to their user groups. The schedulers will schedule users' application instances (both malicious and non-malicious) solely according to resource requirements and other specifications. No profiling or behavior analysis about the application will be initiated to detect malicious activities. To achieve the cost-efficiency requirement of the cloud provider in order to be able to generate revenue, application instances from different users can be co-located on the same physical nodes, and containers/VMs provide required isolation mechanisms. Cluster label-value choices are available for all users to provide affinity specifications.

All users in this paper are considered to have access only to their submitted instances and allocated resources, without any privilege to modify the cluster or monitor cluster information. We consider two types of users, namely normal users and malicious users. Normal users submit tasks without malicious intent to fulfill a computational need. Victim application instances are selected from instances submitted by normal users. We assume malicious users have neither collusion with cloud providers nor any privileges above other users. They do not know the scheduling results of application instances of other users. The only operation they can perform is submitting applications to the cloud like a normal user. But we assume the attackers have information about their victims and can hence generate related specifications by guesses or directly referring to the information they have. Since a lot of cloud users run existing commercial or open-source software (database, data analytics app, etc.) in the cloud to process sensitive data, accessing the target victim software and determining a suitable cloud setup for the software shouldn't be a problem. Also, because the next step of co-location is issuing micro-architectural attacks, which also require analyzing target software, this assumption is reasonable.

Deploying applications to the cloud comes with costs. We assume that users are charged according to their assigned resources and the time they run their applications. In this paper, since we focus more on the attacker side and we only evaluate the cost qualitatively, the cost can be further simplified to be proportional to the number of instances launched on the cloud.

Though there will be no performance profiling and complicated behavioral analysis for applications, obviously suspicious user behaviors like submitting a large number of jobs during a small time frame will be noticed and stopped, which renders brute-force attacks infeasible.

\section{Methodology}\label{SecMethod}

In this section, we will first describe the targeted scheduler in this work. Afterward, we present the proposed co-location attack ({\proj}) in detail. In particular, {\proj} exploits affinity features to locate victims. Hence, co-location can be achieved more efficiently, i.e. with fewer costs.

\subsection{Targeted Scheduler}\label{SecMethodTarget}
\begin{figure*}
\centering
\includegraphics[width=.8\linewidth]{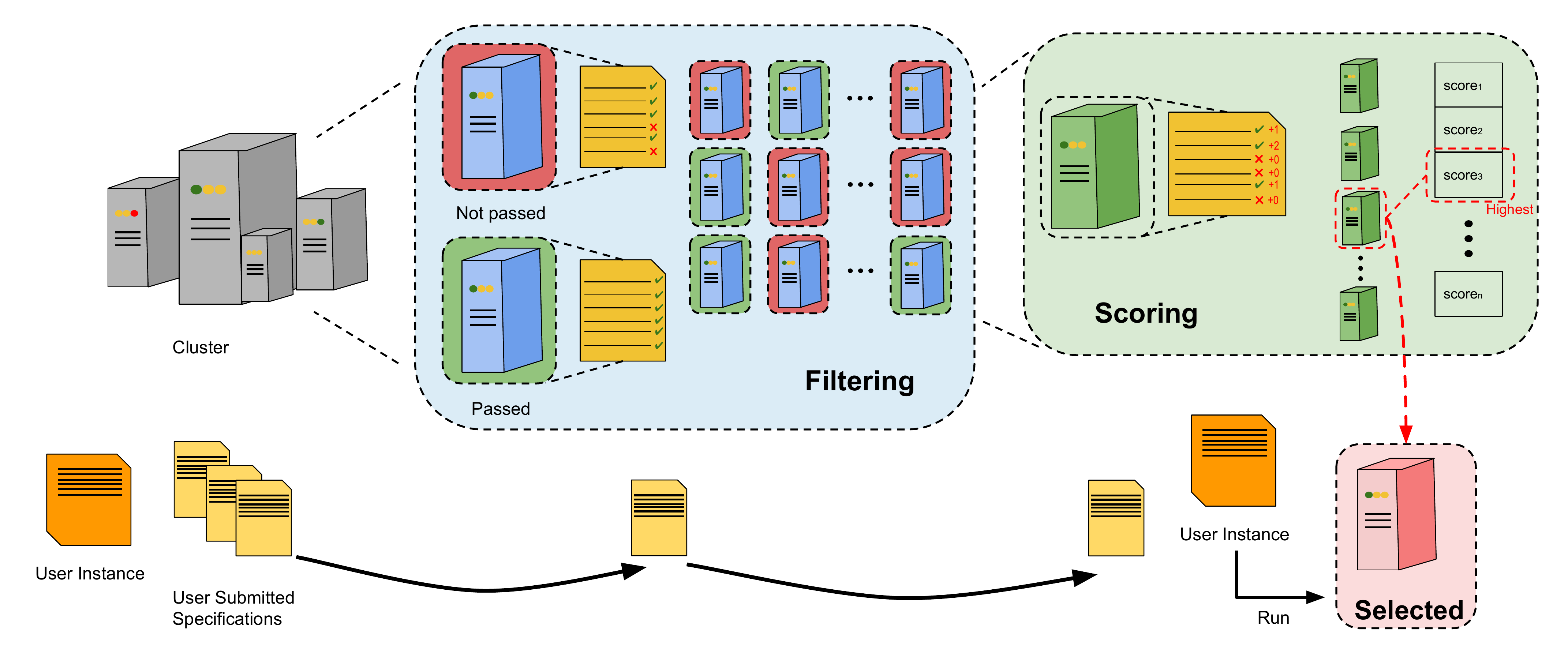}
\caption{The diagram of the scheduler we target. 
%\hl{This diagram should enhance. Lots of white spaces in the figure that does not convey special point! Yo can compact the figure.}
}\label{FigScheduler}
\end{figure*}

In this paper, we target filter-score schedulers (shown in Figure \ref{FigScheduler}), which are widely deployed by various cloud orchestration systems including OpenStack~\cite{openstack}, Kubernetes~\cite{kubernetes} and OpenNebula~\cite{opennebula}. In general, such schedulers are composed of two phases:
\begin{enumerate}
	\item Filtering: in this step, the scheduler excludes nodes that do not have (enough) required resources or do not match other specifications. After this phase, a list of shortlisted nodes will be passed to the scoring phase.
	\item Scoring: in this step, the scheduler tries to find the most suitable node in the list of candidate nodes after the filtering step. It scores each node according to the number of resources available in the node, whether it matches preferences specified by users, and eventually picks the node with the highest score.
\end{enumerate}
The corresponding pseudo-code of this scheduling process is provided in Algorithm \ref{AlgScheduling}.
\begin{algorithm}[htbp]
\caption{Scheduling.}\label{AlgScheduling}
\begin{algorithmic}
\Require User specifications of the application $userspecs$, list of cluster nodes $allnodes$.
\State $candidates$=\textsc{Filter($userspecs$, $allnodes$)}
\State $bestnode$=\textsc{Score($userspecs$, $candidates$)}
\State \Return $bestnode$

\Function{Filter}{$userspecs$, $allnodes$}
\State $candidates$ = $allnodes$
\For{$node$ in $allnodes$}
	\For{$spec$ in $userspecs$}
			\If{$spec$ is not satisfied on $node$}
				\State Delete $node$ from $candidates$
			\EndIf
	\EndFor
\EndFor
\State \Return $candidates$
\EndFunction

\Function{Score}{$userspecs$, $candidates$}
\State $score_{node}=0$ for all $node$ in $candidates$
\For{$node$ in $candidates$}
	\For{$spec$ in $userspecs$}
			\If{$spec$ is resource specification}
				\State Calculate score and add to $score_{node}$
			\ElsIf{$spec$ is satisfied on $node$}
				\State Add a fixed value to $score_{node}$
			\EndIf
	\EndFor
\EndFor
\State \Return $bestnode=\underset{node}{\mathrm{argmax}}    \quad score_{node}$
\EndFunction
\end{algorithmic}
\end{algorithm}

\begin{figure}
    \centering
    \includegraphics[width=.8\linewidth]{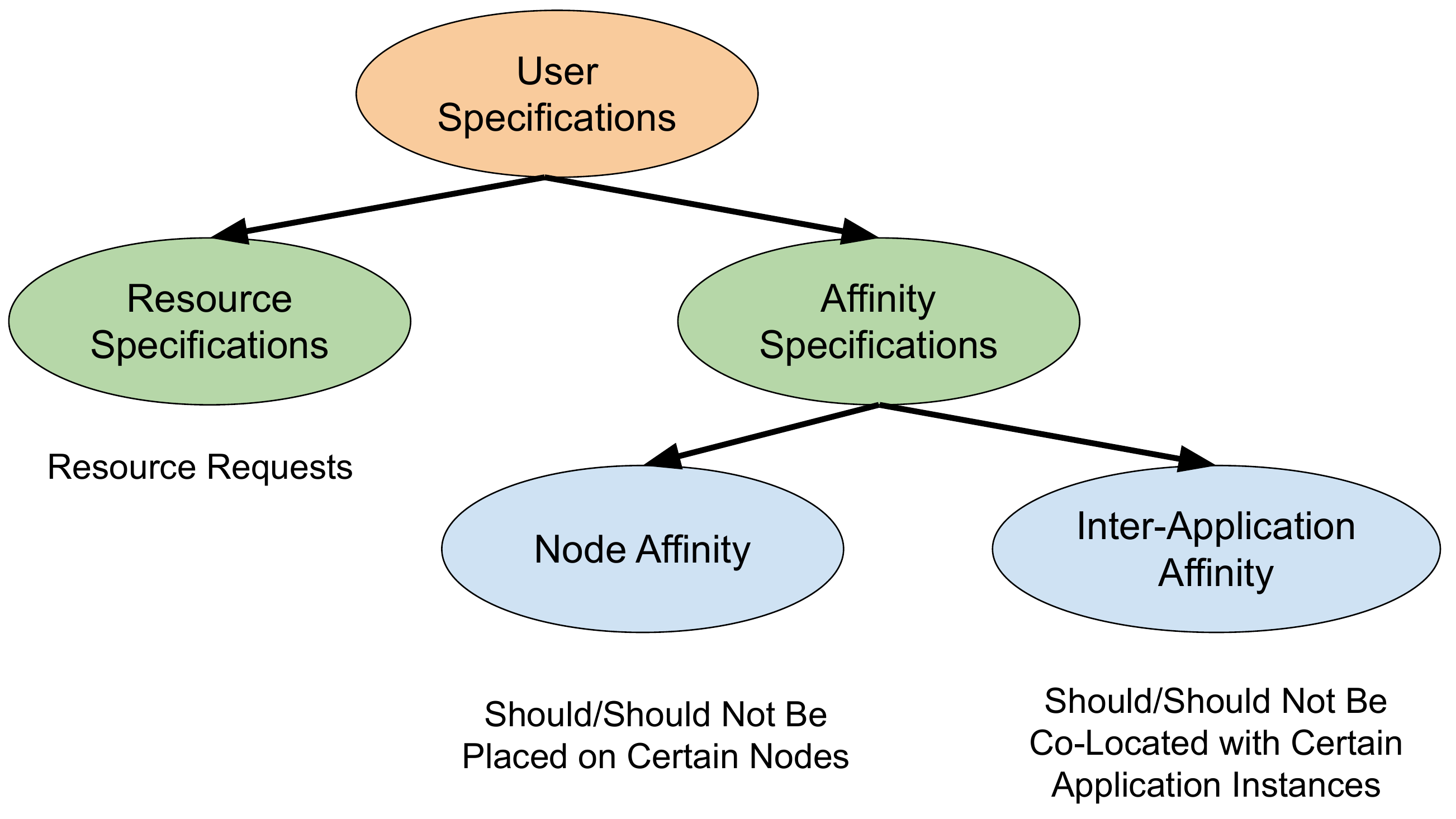}
    \caption{User specifications.}
    \label{FigUserSpecification}
\end{figure}

We conceptually classify user specifications into two categories (as shown in Figure \ref{FigUserSpecification}): resource specification and affinity specifications~\cite{moreno2018orchestrating}. Resource specifications refer to user-specified requirements for resources, including CPU cores, memory space, disk space, etc. On the other hand,  affinity specifications \cite{gulati2012vmware} refer to user-specified requirements or preferences on nodes in the cluster. There are two types of affinity specifications: node affinity/anti-affinity and inter-application affinity/anti-affinity. By specifying node affinity/anti-affinity, users can force the scheduler to schedule an application to/not to nodes with specific resources or features. By specifying inter-application affinity/anti-affinity, users can choose to co-locate/not co-locate with certain types of applications. In general, all affinity features rely on the label system of the cluster and are specified in the format of label-value pairs. Label-value pairs can be added to the metadata information of nodes and application instances to identify different categories of nodes and applications. For example, labels can be added to nodes to indicate the regions the servers belong to or specific hardware resources available on the nodes. In this work, we focus on the latter type of user specifications, and we argue that the extensive use of such features may result in security problems. Those features are used in some scenarios nowadays, including IOT~\cite{santos2019towards}. Also, similar features appear in other commonly used schedulers like SLURM~\cite{yoo2003slurm}. We believe such security problems naturally exist in systems with similar attributes.

\subsection{Attack Strategy}
As we described before, there are features in scheduling algorithms that allow users to submit additional requirements and preferences of nodes and applications to be co-located, enabling users to influence the scheduling results. This exposes vulnerabilities and eases the process of achieving co-location, as attackers can have educated guesses on these feature specifications based on the targeted application and mimic these specifications to increase the probability of co-location.

The intuition behind our strategy is as follows: when two tasks share the same requirements and preferences for nodes and applications running on the nodes, replicating requirements given by victim users helps narrow down the search space in the filtering phase hence the search spaces of the scheduling algorithm will be reduced to similar lists of nodes. Replicating preferences helps to force the target node to get the highest score, increasing the co-location probability, especially in a heterogeneous cluster. As long as the number of resource requirements of attack instances can be set to a minimum, contention can be decreased to the lowest level. The attacker can issue multiple attacks to increase the co-location rate he can achieve and utilize anti-affinity features to spread all the attack instances. This is feasible because such schedulers (e.g., Kubernetes scheduler) allow user-defined specified application labels to participate in the scheduling process. Attackers can hence specify dedicated application labels that are only used by themselves to spread the attack instances.

The attack method we propose in this paper is called {\proj}. An attacker can start by checking the document of the target scheduler, look at the scheduling requirements they can submit, and identify features similar to affinity rules~\cite{moreno2018orchestrating,gulati2012vmware} that allow users to specify requirements or preferences on nodes or co-located applications. We assume that the attacker has enough information about the victim application hence can have relatively accurate guesses on the specifications the victim users might provide (regions of nodes, specific hardware resources on nodes, etc.). The attacker then replicates such specifications, limits the resource requirements to the minimum amount required to issue an attack (so that target nodes in the cluster won't be mistakenly filtered out because of resource limitations), and use inter-application anti-affinity (inter-pod anti-affinity in Kubernetes) to spread the multiple attack instances on different machines. The corresponding pseudo-code is shown in Algorithm \ref{AlgAttack}.

\begin{figure*}
\hfill
     \begin{subfigure}[t]{0.45\textwidth}
         \centering
         \includegraphics[width=\textwidth]{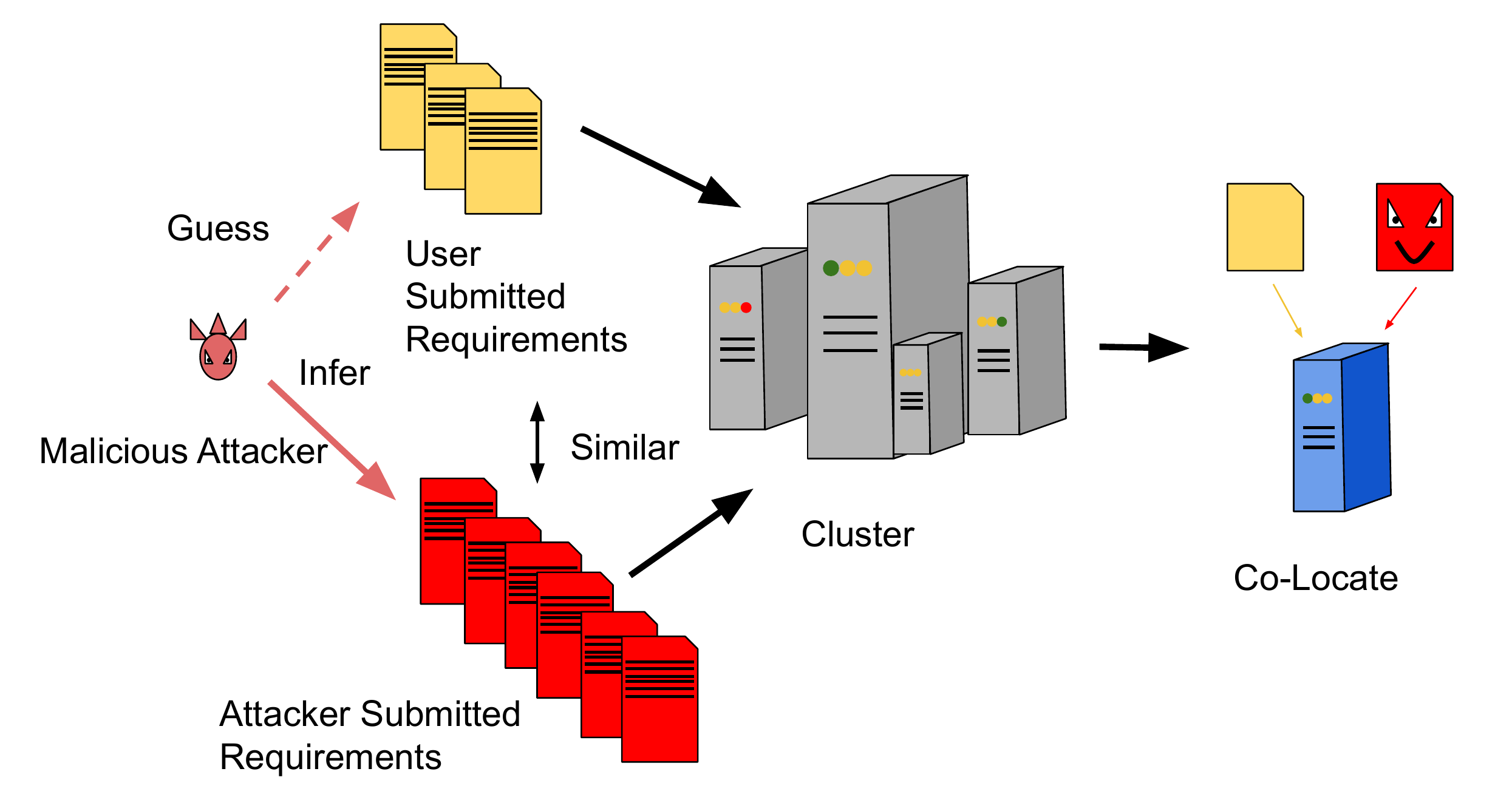}
         \caption{}
        %  \label{FigInstNum:b}
     \end{subfigure}
\hfill
     \begin{subfigure}[t]{0.45\textwidth}
         \centering
         \includegraphics[width=\textwidth]{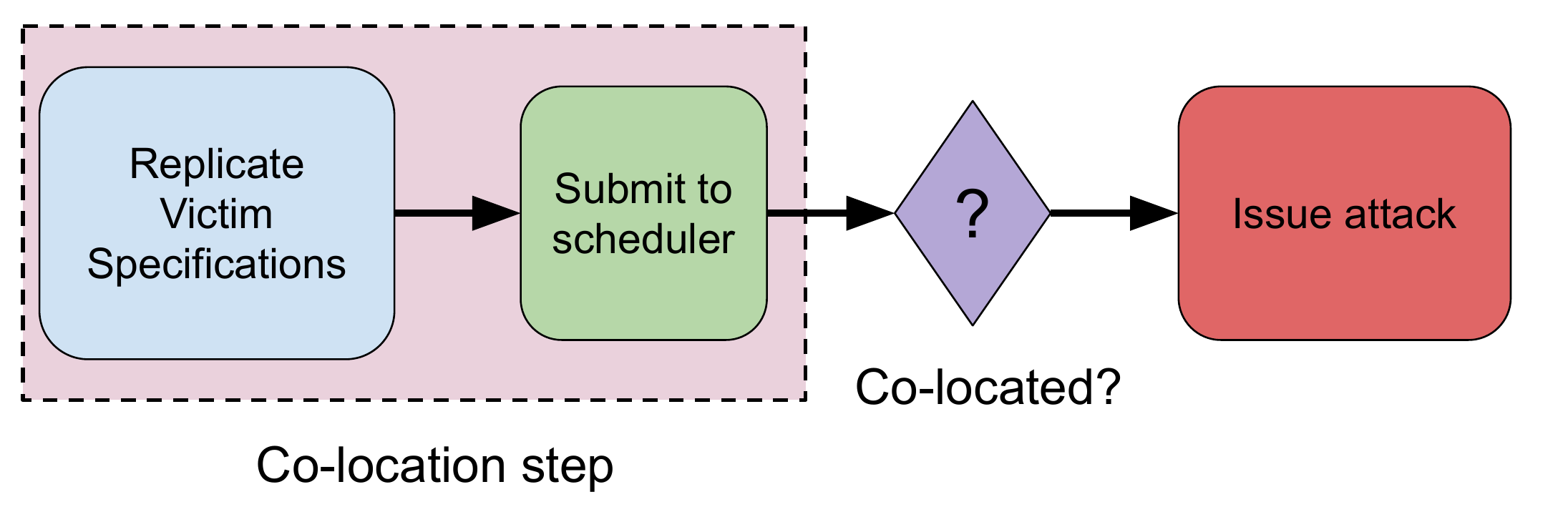}
         \caption{}
        %  \label{FigInstNum:b}
     \end{subfigure}
\caption{Diagram of our attack strategy.}\label{FigAttackMethod}
\end{figure*}

\begin{algorithm}[htbp]
\caption{{\proj}.}\label{AlgAttack}
\begin{algorithmic}
\Require User specifications of the victim application $victimspecs$.
\State $attackspecs$ = \{\}
\For{$spec$ in $victimspecs$}
    \If{$spec$ is an affinity specification}
        \State $attackspecs$ = $attackspecs \cup \{spec\}$
    \EndIf
\EndFor
\State Set all resource requirements to minimum then add to $attackspecs$.\\
\If{There are multiple attack instances}
    \State Add an anti-affinity spec to $attackspecs$:\\
        \hspace{1cm}Label = $\kappa$\\
        \hspace{1cm}Value = $\nu$\\
        \hspace{1cm}($\kappa$ and $\nu$ are used only by attackers)
\EndIf
\\
\Return $attackspecs$
\end{algorithmic}
\end{algorithm}

Generally, suppose an attacker cannot directly obtain the user-submitted specifications. In that case, the attacker can start by studying the target victim application and determine whether there are any special requirements (e.g., victim application requires extensive GPU resources and requires not being co-located with a specific type of application because of contention, etc.). Location preferences of servers of other users can also be exploited, e.g., in~\cite{santos2019towards} round-trip time information is used to specify labels of machines and used as an affinity feature in a Kubernetes-based cluster. 

To increase the co-location success rate, attackers can have multiple versions of attack instances consisting of various user specifications. Compared to brute-force attacks, the cost is still controllable. %For simplicity, in this paper we assume the attackers can produce accurate predictions.

\section{Evaluation}\label{SecExperiments}
Our experiments consist of behavioral simulation and tests in a real cluster. Due to our limited access to large clusters, part of the experiments are conducted in a scheduler simulator implemented in Python that simulates the behaviors of filter-score schedulers. The simulator gives us the flexibility to test the influence of different parameters. The test on a real cluster is conducted on a public cloud testing environment in which we deploy a $40$-node Kubernetes~\cite{kubernetes} cluster and randomly submit jobs to observe the scheduler outcome of a real cluster. In our experiments, we only consider whether or not the target applications can be co-located. 

\subsection{Experiments in Simulation}
\subsubsection{Experiment Settings}
The simulator written in Python consists of approximately $1000$ LOC. In the simulation process, configurations of machines in the cluster are generated randomly. There are three phases in the simulation process: node setup, application generation, and execution simulation. 

For node settings, we specify the resource capacities (number of CPU cores, memory space, storage space, network ports) of every node, as well as the labels of each node. In our simulation, the number of nodes in the cluster is $100$. In the execution phase, we split the whole experiment into $1000$ slots (scheduling cycles). In each scheduling cycle, we submit $10$ random applications to the scheduling queue. The total number of labels used in our experiments is $11$, including $5$ as node labels, $5$ as application labels, and $1$ used as spreading labels for attackers to spread attack instances in the cluster. 

Application requirements submitted to the system are randomly generated by the simulator. Besides required resources to run the application, labels of the applications and scheduling requirements and preferences options are generated randomly. When generating labels for nodes/applications, we follow the following procedure. Assume there are $n$ labels for nodes/applications. We allow some labels to be missing in the configuration, and the probability of having a single label is $p_m$. To generate all label specifications, we examine the $n$ labels one by one, and for each label, with a probability $p_m$ we randomly assign a value to that label. Otherwise, we set the label as missing. The process of generating affinity specifications is similar.

After that, we enter the execution phase, where we submit various applications, including normal tasks and malicious tasks that aim to accomplish co-location with some of the normal tasks. The execution phase is implemented synchronously, meaning that we divide the whole experiment into a large number of slots and will push several applications to the scheduling queue, make scheduling decisions and collect scheduling results in each time slot. Thousands of tasks will be provided to the scheduler, including victim tasks, attack instances, and randomly generated unrelated applications that aim to mimic the assumed unpredictable user behaviors in the system. The goal of each malicious application instance is to co-locate with its target victim instance. We repeat this attack during the execution phase and target different victims each time. After the execution phase, we calculate the co-location rate and provide corresponding results.

\subsubsection{Influence of User Specifications}
\begin{figure*}[ht]
\centering
     \begin{subfigure}[b]{0.3\textwidth}
         \centering
         \includegraphics[width=\textwidth]{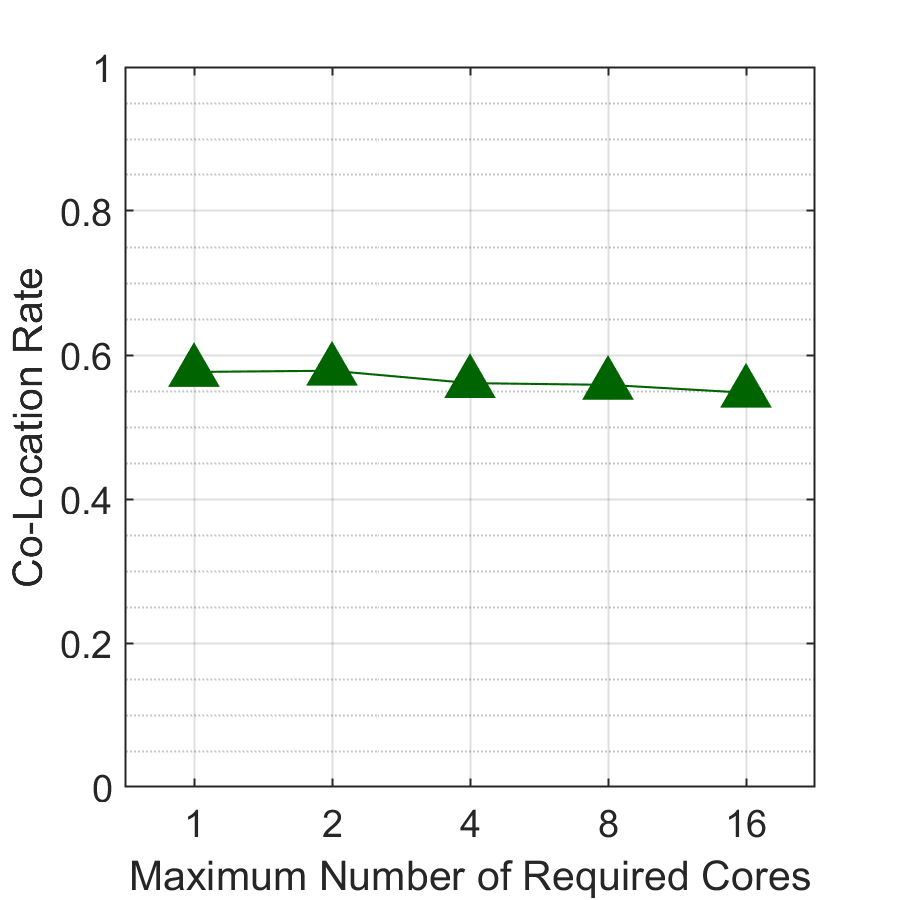}
         \caption{Varying the number of required CPU cores.}
         \label{FigVaryRes:a}
     \end{subfigure}
     \hfill
     \begin{subfigure}[b]{0.3\textwidth}
         \centering
         \includegraphics[width=\textwidth]{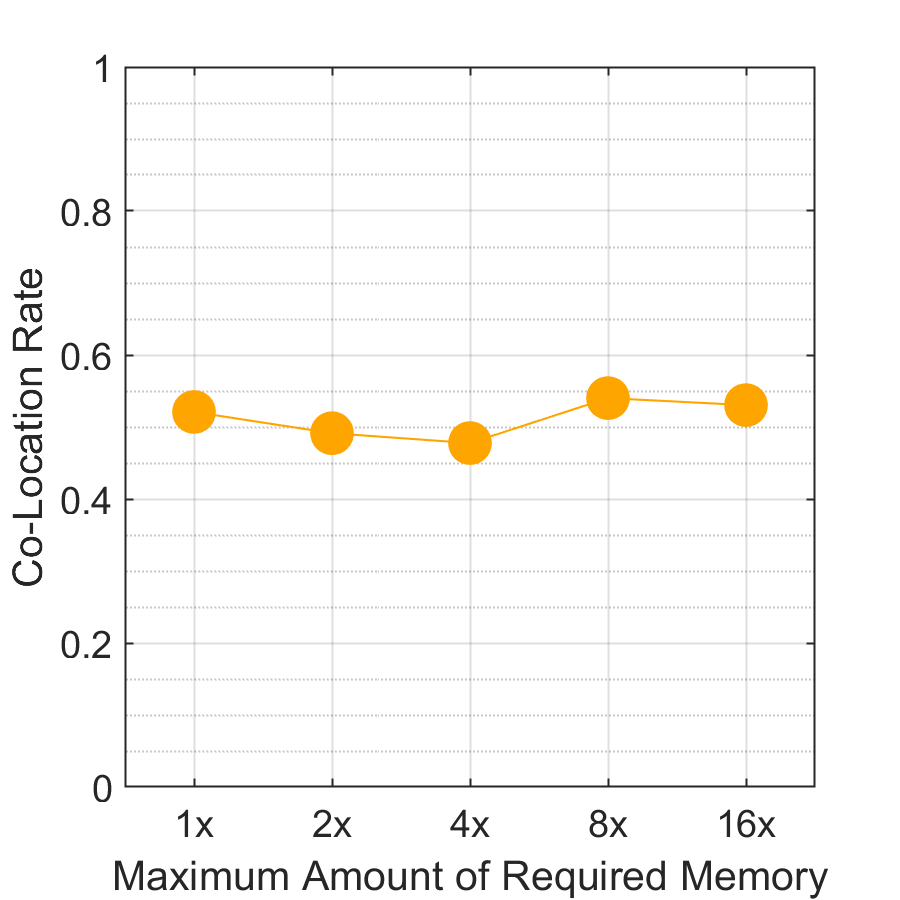}
         \caption{Varying the amount of required memory.}
         \label{FigVaryRes:b}
     \end{subfigure}
     \hfill
     \begin{subfigure}[b]{0.3\textwidth}
         \centering
         \includegraphics[width=\textwidth]{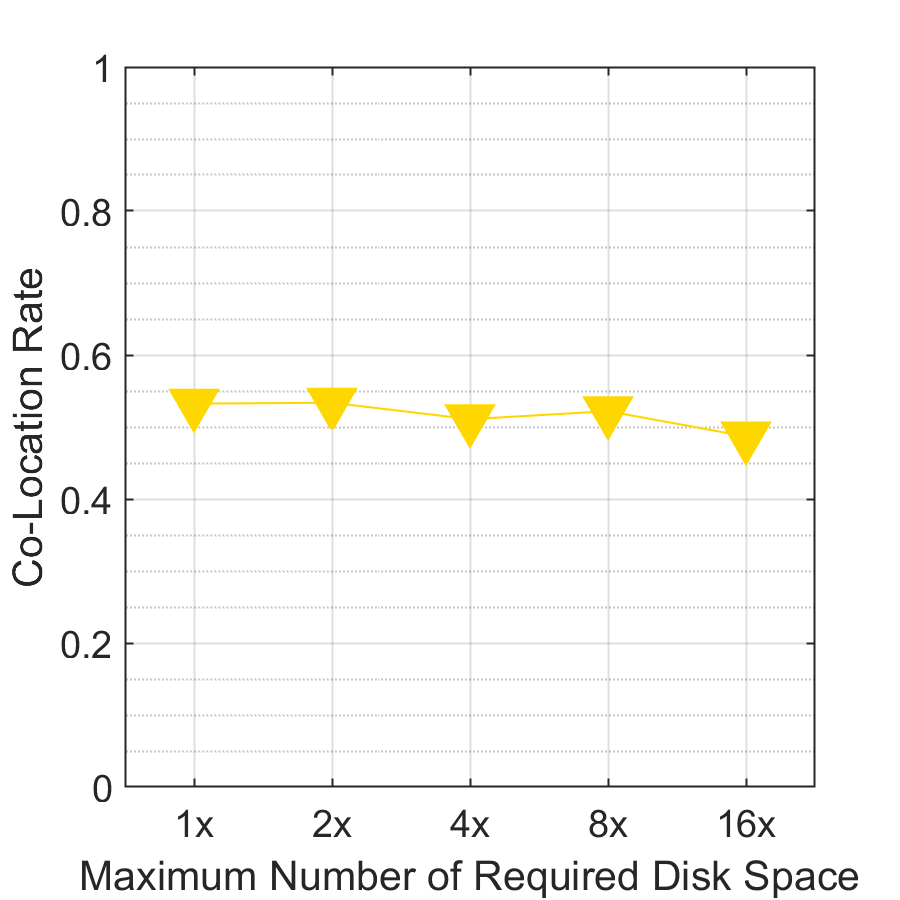}
         \caption{Varying the amount of required disk space.}
         \label{FigVaryRes:c}
     \end{subfigure}
        \caption{The influence of resource specifications.}
        \label{FigVaryRes}
\end{figure*}

We first examine how resource specifications can affect the co-location rate of {\proj}. We evaluate the co-location rate under different amounts of resources (CPU cores, memory, disk space) that users can request and provide the results in Figure \ref{FigVaryRes}. Figure \ref{FigVaryRes:a}, Figure \ref{FigVaryRes:b} and Figure \ref{FigVaryRes:c} show the co-location rate under different maximum limitation settings for the number of CPU cores, amount of memory space, and amount of disk space respectively where $1$x$=512$MB for memory and $1$x$=16$MB for disk. The amount of user-requested resources follow a uniform distribution between $0$ and the given maximum value. Though some scheduling algorithms (e.g., Kubernetes~\cite{kubernetes}) do not directly check disk space, we still do that in our simulations in case future scheduler designs take disk space into consideration. As shown in all results, these resource specifications will not influence the co-location rate, as the required resources change $16$x the change in co-location rate is no more than $5\%$. Therefore we will focus on affinity features in our later experiments.

% we change the maximum number of CPU cores a user can request. In , we change the maximum amount of memory space a user can request. Here, $1$x$=512$MB. In Figure \ref{FigVaryRes:c}, we change the maximum amount of disk space. Here, $1$x$=16$MB. The amount of user-requested resources follow a uniform distribution between $0$ and the given maximum value. Though some scheduling algorithms (e.g., Kubernetes~\cite{kubernetes}) do not directly check disk space, we still do that in our simulations. As shown in all results, these resource specifications will not influence the co-location rate, as the required resources change $16$x the change in co-location rate is no more than $5\%$. Therefore we will focus on affinity features in our later experiments.

\begin{figure}[htbp]
\centering
         \includegraphics[width=\linewidth]{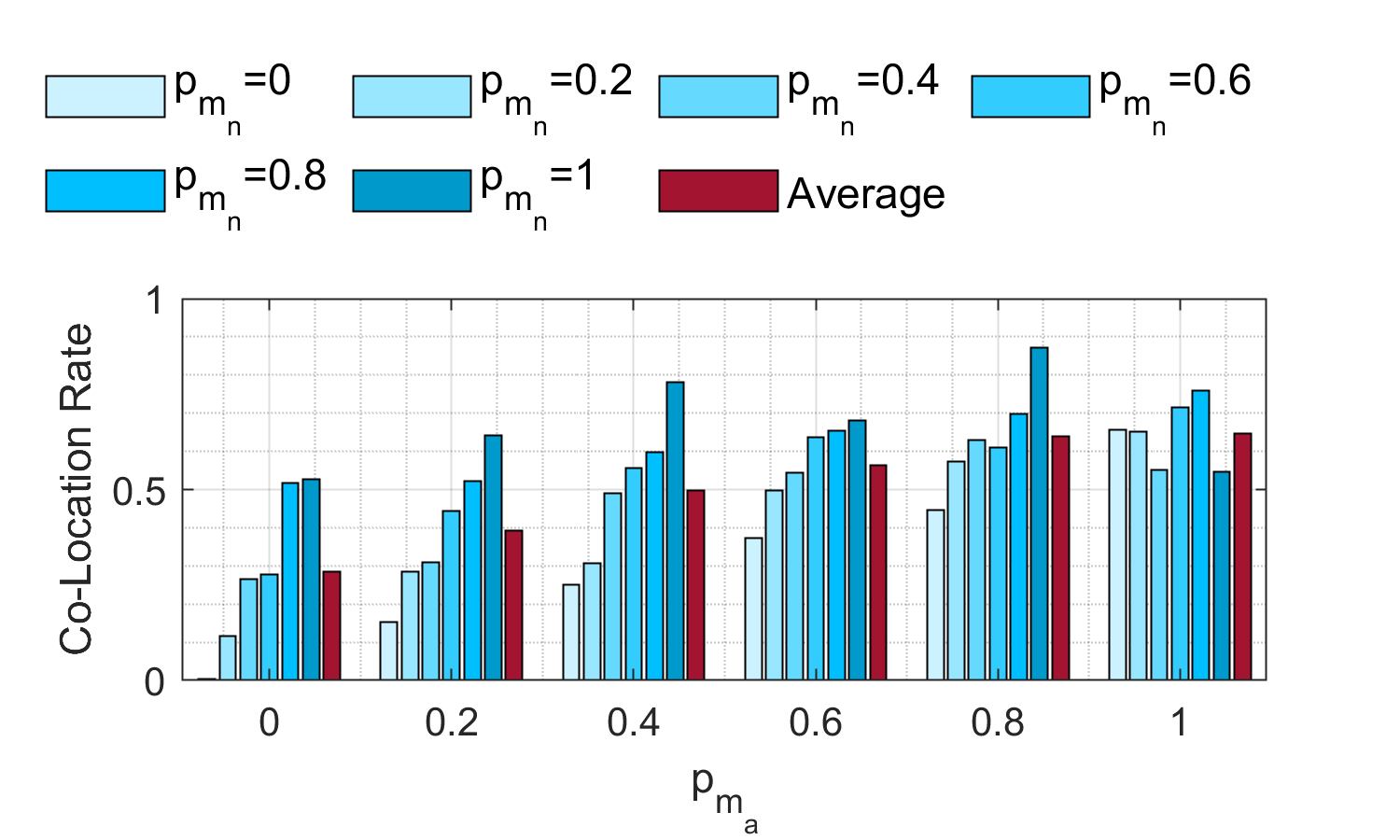}
         \caption{Influence of node-affinity and inter-application affinity features.}
         \label{FigAffinity}
\end{figure}

Figure \ref{FigAffinity} presents the influence of specifications of node affinity/anti-affinity as well as inter-application affinity/anti-affinity on co-location rate. In this experiment, we only issue one attack instance each time. As we mentioned before, we use the probability $p_m (0 \leq p_m \leq 1)$ to control the number of affinity features used in the experiment. When we have $n$ labels for specifying, the mean number of label specifications that will be considered during scheduling is $np_m$. We denote the $p_m$ used to generate node affinity/anti-affinity specifications as $p_{m_n}$ and the $p_m$ used to generate inter-application affinity/anti-affinity specifications as $p_{m_a}$. Increasing the number of node-affinity/anti-affinity and inter-application affinity/anti-affinity specifications submitted by users will both raise the co-location rate malicious attackers can achieve. As shown in Figure \ref{FigAffinity}, within each group of bars which share the same $p_{m_n}$, increasing $p_{m_a}$ generally increases the co-location rate. Comparing bars with the same $p_{m_n}$ for node-affinity/anti-affinity features, we have the same observation regarding the influence of $p_{m_a}$. Despite some exceptions caused by randomness in the experiments, the trend well matches our expectation that adding more affinity specifications to control the output of schedulers will narrow down the search space of nodes and cause the scheduling results more foreseeable, which attackers can exploit to achieve co-location.

\subsubsection{Influence of Number of Attack Instances}\label{SubSecNumAttackSim}
Instead of launching only one attack instance at a time, malicious users can submit multiple instances to the scheduler to improve the co-location rate. Previously, in~\cite{ristenpart2009hey} the authors proposed to use a brute force strategy to spread their attack program across the cloud and achieve co-location. Han \textit{et.al}~\cite{han2015using} further formalizes the problem and proves the lower bound of the number of VMs that should be submitted to achieve co-location. However, submitting a substantial number of applications to the scheduler within a small time slot will be easy to detect, and the cost will be unacceptable in some circumstances. We present in this part that, by using {\proj}, malicious users can achieve a relatively high co-location rate among around $4000$ victim instances in a $100$-node cluster with affinity features used by users.

\begin{figure}[htbp]
    \centering
    \includegraphics[width=.9\linewidth]{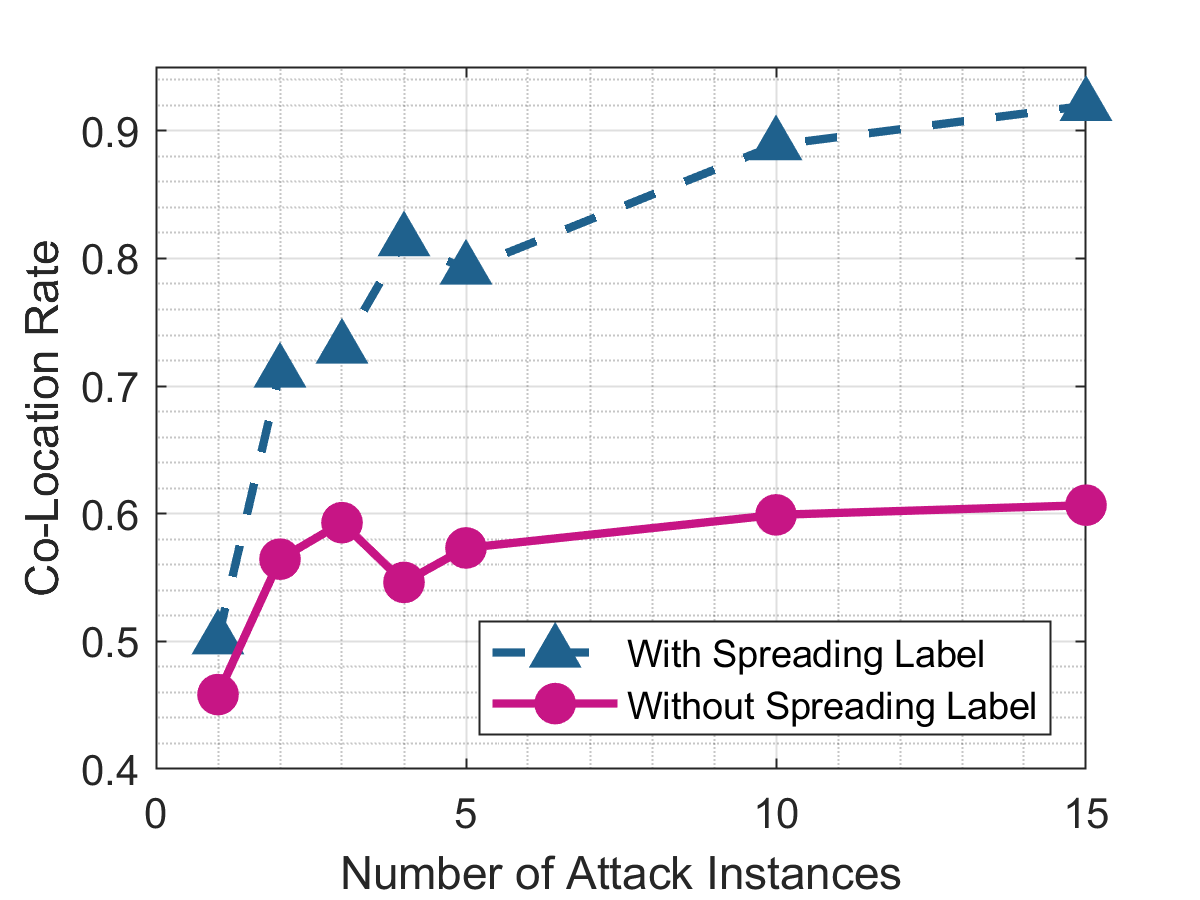}
    \caption{The influence of the number of issued attack instance.}
        \label{FigInstNum}
\end{figure}

Figure \ref{FigInstNum} shows the simulation results of co-location rates that a malicious attacker can achieve by issuing different numbers of attack instances. Firstly, we directly submit multiple attack instances to the scheduler following the strategy in Section \ref{SecMethod} but without utilizing the inter-application anti-affinity to spread all attack instances. The result (the red line in Figure \ref{FigInstNum}) shows that the co-location rate does not improve much, which matches our expectation. The reason is that since multiple attack instances may be co-located on the same node that the victim does not reside in hence will not show significant advantages over the single-issue attack. Next, we demonstrate how involving anti-affinity features to spread attack instances improves the co-location rate. We reserve a special label for the use of attackers in our simulation called spreading label and specify inter-application anti-affinity to prevent attack instances to locate together. As shown in the blue line in Figure \ref{FigInstNum}, the co-location rate grows as the number of attack instances increases, and with more than $1$ attack instances issued, the co-location rate reaches around $20\% - 30\%$ higher than without using a spreading label. 
% \begin{figure*}[htbp]
% \centering
%      \begin{subfigure}[b]{0.45\textwidth}
%          \centering
%          \includegraphics[width=\textwidth]{Figures/FigVaryNum.png}
%          \caption{Directly issuing multiple attack instances. \hl{Please change the chart to line as it shows a trend!}}
%          \label{FigInstNum:a}
%      \end{subfigure}
%      \hfill
%      \begin{subfigure}[b]{0.45\textwidth}
%          \centering
%          \includegraphics[width=\textwidth]{Figures/FigVaryNumAnti.png}
%          \caption{Using anti-affinity features to spread attacks across the cluster. \hl{Please change the chart to line as it shows a trend!}}
%          \label{FigInstNum:b}
%      \end{subfigure}
%         \caption{The influence of the number of issued attack instance.}
%         \label{FigInstNum}
% \end{figure*}

\subsection{Experiments in Cluster}
\subsubsection{Experiment Settings}
To validate the results of our simulator, we also conduct the experiments in a real cloud. We test related features in a Kubernetes~\cite{kubernetes} cluster deployed on CloudLab~\cite{duplyakin2019design}. CloudLab provides cloud researchers control and access to real cluster hardware, even down to bare-metal level. Since we run experiments on top of the Kubernetes cluster setup on CloudLab’s hardware and only interact with Kubernetes, we believe our settings are realistic. Also, since there are other cloud schedulers using algorithms similar to Kubernetes, and we do not rely on certain attributes of Kubernetes, the results collected on this cluster are representative.

CloudLab offers users dedicated ready-to-run clusters, with all required settings contained in a cluster profile. Kubernetes uses a filter-score scheduling strategy and is susceptible to the manipulation of scheduling results by submitting some affinity requirements or preferences. We deploy a cluster with \textit{k8s} profile and directly interact with the cluster using Kubernetes command-line tool \texttt{kubectl}. The cluster in which we deploy our experiments consists of $40$ nodes. At first, we use the command \texttt{kubectl label nodes <nodename> <label>=<value>} to label each machine with some randomly generated configuration information. These configurations include CPU type, GPU type, disk type, and memory capacity ('high' or 'low' similar to~\cite{santos2019towards}). After that, we randomly generate \texttt{.yaml} files describing different pods (in Kubernetes, user instances are encapsulated in containers and deployed in the unit of pods, which normally contain one container per pod). The container type of each non-malicious job is randomly chosen from popularly used containers from Docker Hub~\cite{dockerhub}. Node-affinity/anti-affinity as well as inter-pod affinity/anti-affinity are also generated in each \texttt{.yaml} file. After using the command \texttt{kubectl apply -f <jobname.yaml>} to deploy and schedule all pods, we run a script to test whether each target victim-attacker pod pair is co-located and report the co-location rate. We assume attackers target all the separate victims one at a time and issue corresponding attack instances. Launching attack instances to achieve co-location with a victim is defined as an attack. The testing script only considers pods that are successfully deployed and calculates the ratio of successful attacks. In this part, we aim to validate that the trends are the same as shown in simulation results.

\subsubsection{Influence of Affinity Features}\label{SecAffinityCL}
We first examine whether or not simply duplicating user specifications increases the co-location rate in the cloud. In each test of our experiment, we assume all users provide the same number of affinity specifications. We divide all affinity features into four categories: required node affinity/anti-affinity specifications, preferred node affinity/anti-affinity specifications, required pod affinity/anti-affinity specifications, preferred pod affinity/anti-affinity specifications. We use $4$-digit numbers to differentiate the tests, and each digit denotes the number of labels specified in that category. For example, `$2131$' refers to a test where all submitted pods have: $2$ required node affinity/anti-affinity specifications, $1$ preferred node affinity/anti-affinity specification, $3$ required pod affinity/anti-affinity specifications and $1$ preferred pod affinity/anti-affinity specification. In every test of this experiment, we deploy a total number of $800$ randomly generated pods, where $200$ is considered victims. %The results are provided in Figure \ref{FigAffinityCL}.

% In every test of this experiment, we deploy a total number of $800$ randomly generated pods, where $200$ is considered victims. The results are shown in Figure \ref{FigAffinityCL}.

\begin{figure*}
\centering
\includegraphics[width=.8\linewidth]{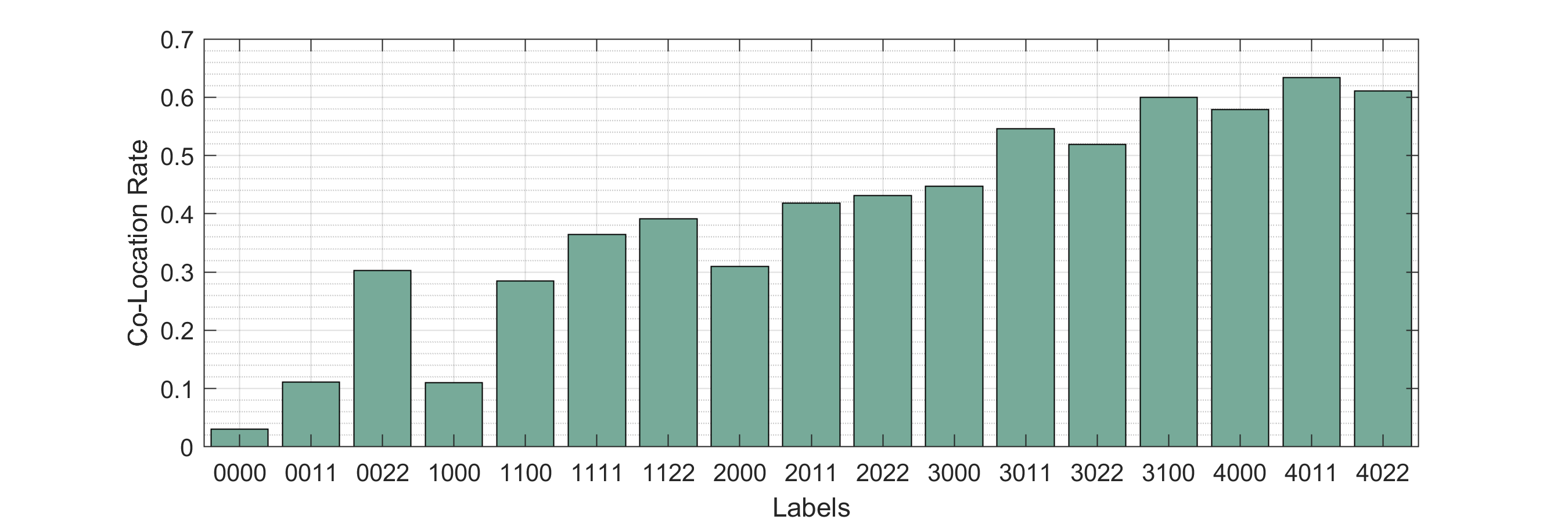}
\caption{The influence of affinity features on co-location rate.}\label{FigAffinityCL}
\end{figure*}

As shown in Figure \ref{FigAffinityCL}, the co-location rate increases as the users use more affinity features. By only issuing a single attack instance, the co-location rate can achieve as high as $60\%$. This matches our results in our simulator, showing that the risk indeed exists in real clusters. This result again showcases how affinity features can be dangerous if attackers exploit them to achieve co-location.

\subsubsection{Issuing Multiple Attack Instances}
\begin{figure*}[htbp]
\begin{subfigure}[t]{0.45\linewidth}
         \centering
         \includegraphics[width=\textwidth]{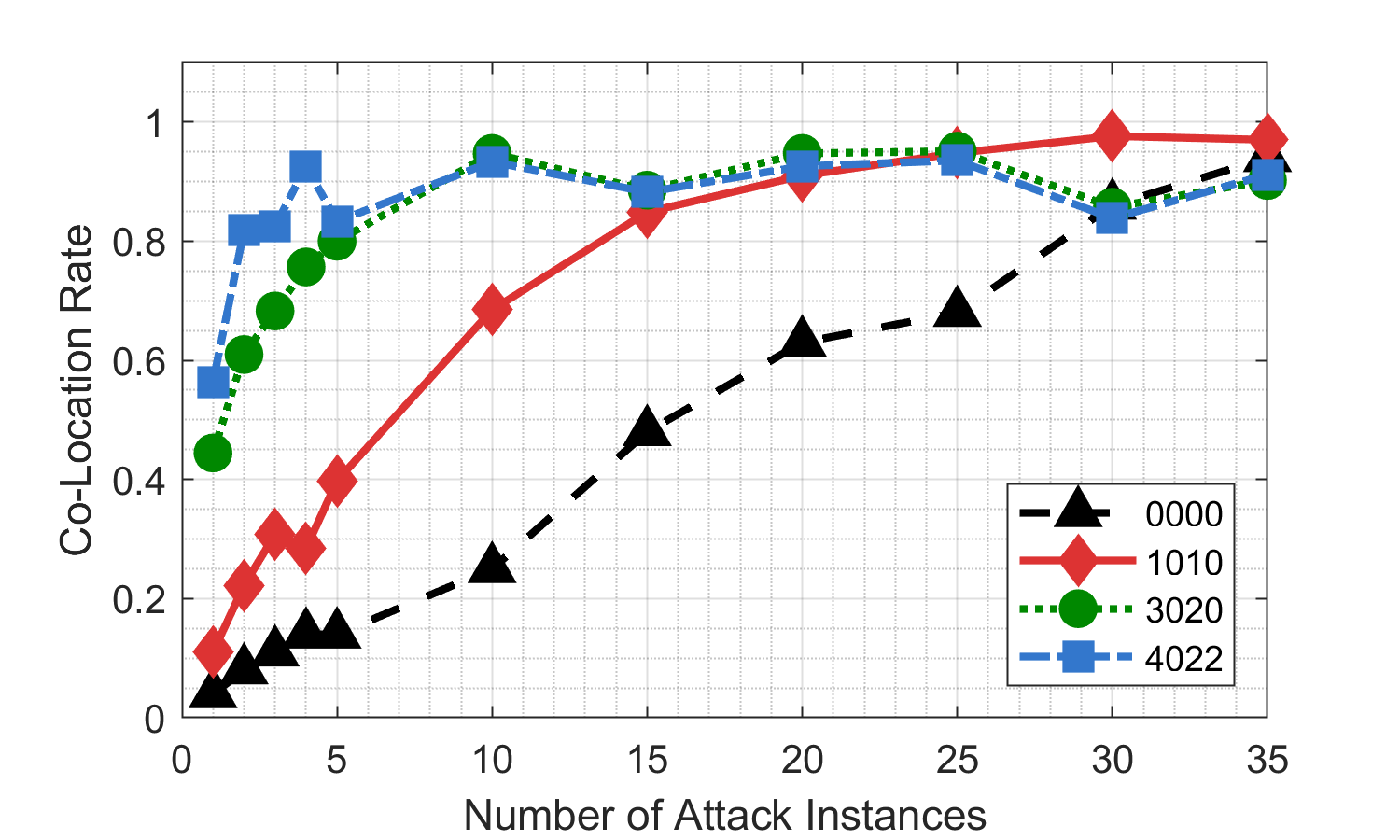}
         \caption{Varying number of affinity features used.}
         \label{FigVaryNumCL:a}
     \end{subfigure}
     \hfill
     \begin{subfigure}[t]{0.45\linewidth}
         \centering
         \includegraphics[width=\textwidth]{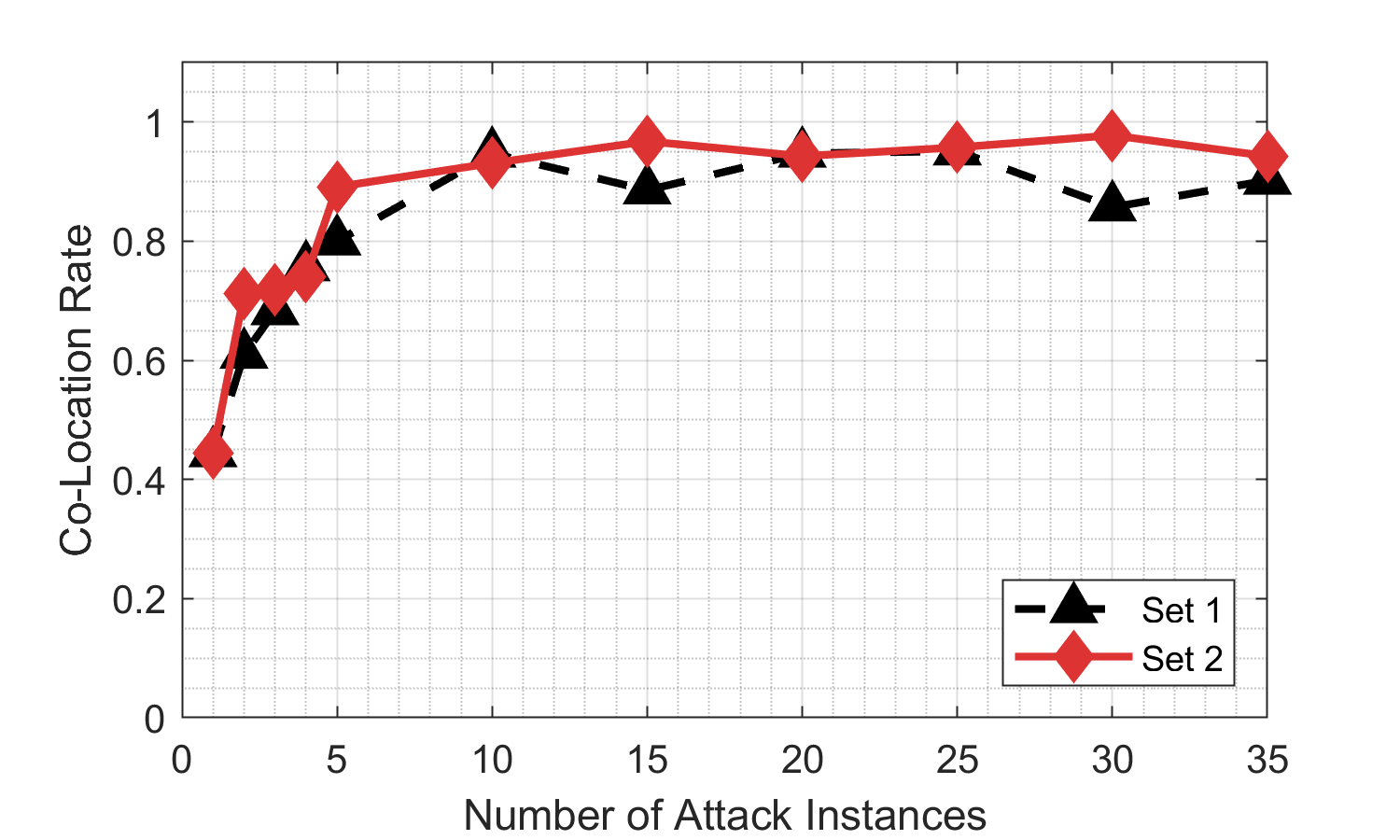}
        \caption{Varying application types.}
        \label{FigVaryNumCL:b}
     \end{subfigure}
     \hfill
     \begin{subfigure}[t]{0.45\linewidth}
         \centering
         \includegraphics[width=\textwidth]{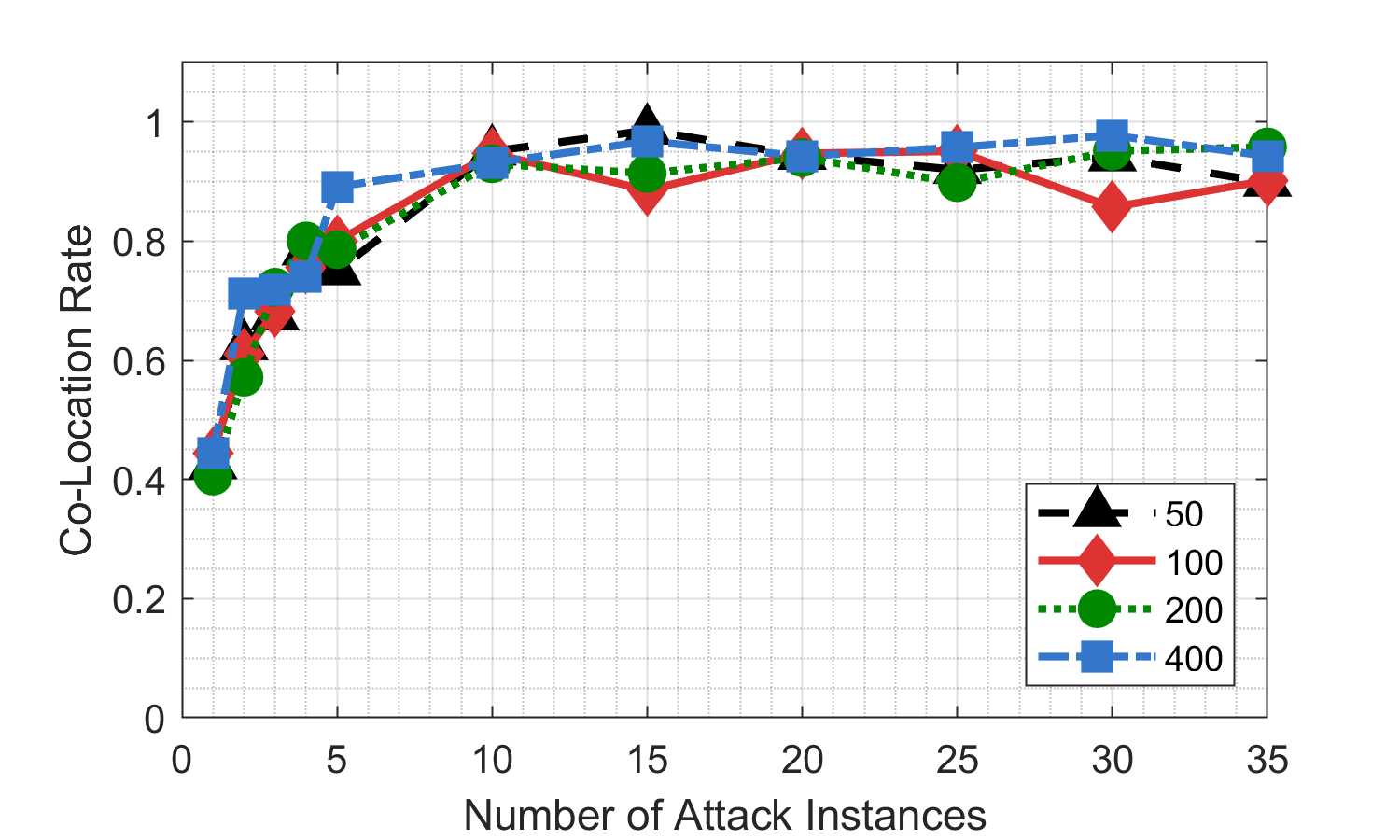}
        \caption{Varying the number of targets.}
        \label{FigVaryNumCL:c}
     \end{subfigure}
     \hfill
     \begin{subfigure}[t]{0.45\linewidth}
         \centering
         \includegraphics[width=\textwidth]{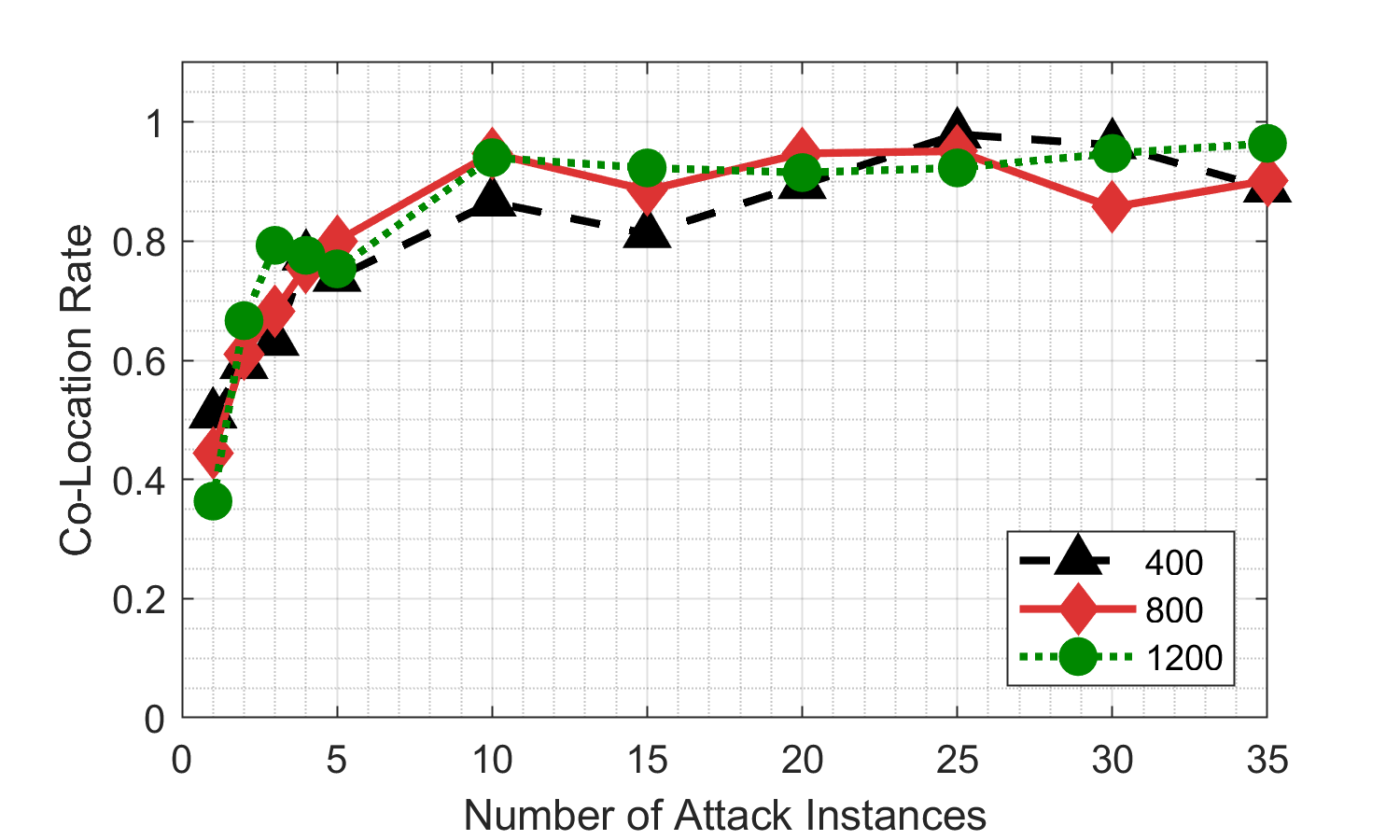}
        \caption{Varying the total number of deployed applications.}
        \label{FigVaryNumCL:d}
     \end{subfigure}
    
         \caption{Influence of number of issued attack instances.}
         \label{FigVaryNumCL}
\end{figure*}

After constructing a Kubernetes cloud in CloudLab platform, we also test the influence of the number of attack instances issued by malicious users. The results are shown in Figure \ref{FigVaryNumCL}.

In this experiment, we vary the number of attack instances a malicious attacker can issue as well as several other settings and deploy all the non-malicious applications and malicious attack instances to the cloud in a single experiment. We assume attackers utilize a special label-value pair to mark their malicious instances and exploit the pod anti-affinity feature to avoid the same attack instances to locate on the same machine. From all results shown in Figure \ref{FigVaryNumCL}, we can see that as the number of attack instances the attacker launches increases, at first, there is a significant gain in co-location rate an attacker can achieve. However, after the number of attack instances reaches a certain value, increasing the number of attack instances does not provide an obvious gain in co-location rate anymore. The curve of the co-location rate shows an interesting roofline pattern.

To further explore what parameters may affect the roofline model, we vary different parameters and show how the co-location rate curve is affected. In Figure \ref{FigVaryNumCL:a}, we vary the number of affinity features used in the way similar to Section \ref{SecAffinityCL}. As shown in Figure \ref{FigVaryNumCL:a}, with more affinity features used by users,
\begin{enumerate}
    \item the initial co-location rate is higher;
    \item the growth of co-location rate as the number of attack instances increases is more rapid;
    \item the saturation point is reached earlier.
\end{enumerate}
It is worth noting that the curve '$0000$' can be considered as the results under a brute-force attack setting, since no affinity features are utilized here. As a result, the initial co-location rate is the lowest, and it costs more instances to reach a similar co-location rate.

In the following $3$ experiments, affinity features are fixed to $3020$ pattern to provide suitable lengths of growing stage and steady stage. In the experiment of Figure \ref{FigVaryNumCL:b}, we test the co-location rate on two sets of containers. Set $1$ consists of Traefik, Ngnix, Tomcat, Redis, Mongo, and Wordpress. Set $2$ consists of Alpine, Busybox, Python, Registry, Httpd, and Memcached. In the experiments of Figure \ref{FigVaryNumCL:c} and \ref{FigVaryNumCL:d} we vary the application types, the number of targets in the experiment, and the total number of applications, respectively. As shown in the results, varying these parameters lead to similar co-location rates under different attack instance numbers. These parameters are hence considered not to influence co-location rate.

The trend can be explained as follows. Since attackers specify anti-affinity requirements when they submit their tasks in order to spread their attack instances across the entire cluster, the increase in co-location rate at first is because more qualified nodes after the filtering step are occupied when the number of attack instances increases. When victims use more affinity features, the search space narrows down; hence the co-location rate will grow faster as there are fewer qualified nodes. The co-location rates reach the saturation stage after the number of attack instances passes a specific number because nearly all possible nodes are occupied. However, the later steady stage also demonstrates that increasing the number of attack instances cannot cover all possible nodes, and there will still be victims escaping from such attacks. This phenomenon can be possibly caused by the variations of cluster states (resource usage in each node, etc.) during different scheduling processes since all specifications of the attack instances are the same as victims. Therefore, when the number of attack instances exceeds a certain number, the co-location rate stays at a relatively high value but will not reach $100\%$.

\section{Mitigation}\label{SecMitigation}
\subsection{Can Migration Defend Against Co-Location Attacks?}
In some previous research works in this area~\cite{zhang2012incentive}, migration of application instances in the cloud is used as a defense method. By moving application instances occasionally, the difficulty of achieving co-location will increase for attackers. However, in this part, we argue that under the assumption that migration still respects user requirements, the risk of being co-located with malicious instances is potentially higher.

Co-location rate in our experiment is defined as
\begin{equation*}
\text{Co-Location Rate} = \dfrac{\text{Successfully Co-Located Attacks}}{\text{Total Number of Attacks}}.
\end{equation*}
The count of successful attacks, however, only considers the initial placement. When migration is introduced, attack instances that are previously considered successful may not co-locate with victim instances anymore, but new co-located attack-victim pairs may appear as well. We hence modify our definition of a successful attack. An attack attempt is considered successful when the target victim application is located on nodes with attack instances in $t\%$ of its lifetime, where $t$ is a predefined threshold percentage. In our experiments, we set it to $80$, so when an application is co-located with an attack instance targeting itself for more than $80\%$ of its lifetime, we consider the corresponding attack a successful one. Attacks like performance degradation attack~\cite{allan2016amplifying,varadarajan2012resource} can benefit from longer co-location time.

\begin{figure}
    \centering
    \includegraphics[width=\linewidth]{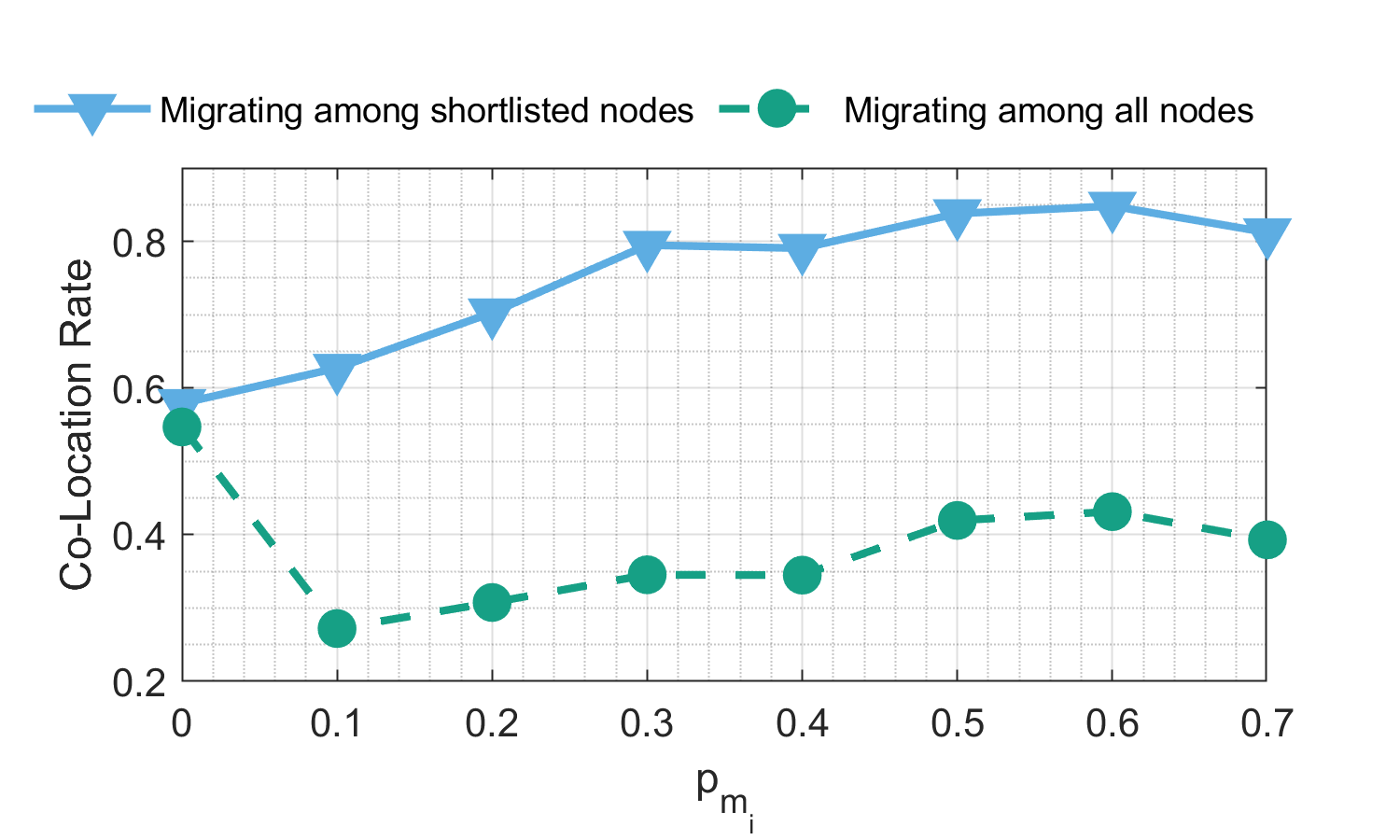}
    \caption{Co-location rate results in our simulator after introducing migration.}
    \label{FigMigration}
\end{figure}

The migration process in our experiment is implemented as follows. At the end of each time slot, we examine every application, and with a probability $p_{m_i} (0 \leq p_{m_i} \leq 1)$, we remove it from its current node and place it on another node. There are two methods to select migration destinations:
\begin{enumerate}
\item Randomly selects one node from the shortlist of nodes after the filtering step.
\item Randomly selects one node from the cluster.
\end{enumerate}
The results are presented in Figure \ref{FigMigration}. In these experiments, we vary $p_{m_i}$ to control how frequently instances in the system are being migrated. The first selection method respects user-specified requirements since the migration destination is chosen from the filtered shortlist. The second selection method brings more randomness, and as a trade-off, the user requirements can be violated shortly after its initial placement.

When the migrating destination selection is based on the shortlist of nodes after the filtering step, as we migrate more frequently, the co-location rate increases, which contradicts our intuition. This can be explained as the result of more collisions between the attack and victim instances in our experiment. This results in the victim instance being co-located with the attacker longer in time, giving rise to more successful attack attempts. When we expand the selection to all nodes in the cluster, there is indeed a drop in the co-location rate. However, we can observe a decrease in defense effect as $p_{mi}$ approaches $50\%$ because of the increase in the probability of collisions between victims and attack instances after adding migrations.

% \begin{figure*}[htbp]
% \centering
%      \begin{subfigure}[t]{0.45\linewidth}
%          \centering
%          \includegraphics[width=\textwidth]{Figures/FigMigration1.png}
%          \caption{Migrate application instances among the shortlisted nodes.}
%          \label{FigMigration:a}
%      \end{subfigure}
%      \hfill
%      \begin{subfigure}[t]{0.45\linewidth}
%          \centering
%          \includegraphics[width=\textwidth]{Figures/FigMigration2.png}
%          \caption{Migrate application instances among all nodes in the cluster.}
%          \label{FigMigration:b}
%      \end{subfigure}
%      \caption{Co-location rate results in our simulator after introducing migration. \hl{These two subfigures can be drawn in one figure as their X and Y axes are the same. Moreover, you can replace them with a line chart!}}\label{FigMigration}
% \end{figure*}

\subsection{Mitigation Strategy}

\begin{figure}
\centering
\includegraphics[width=.85\linewidth]{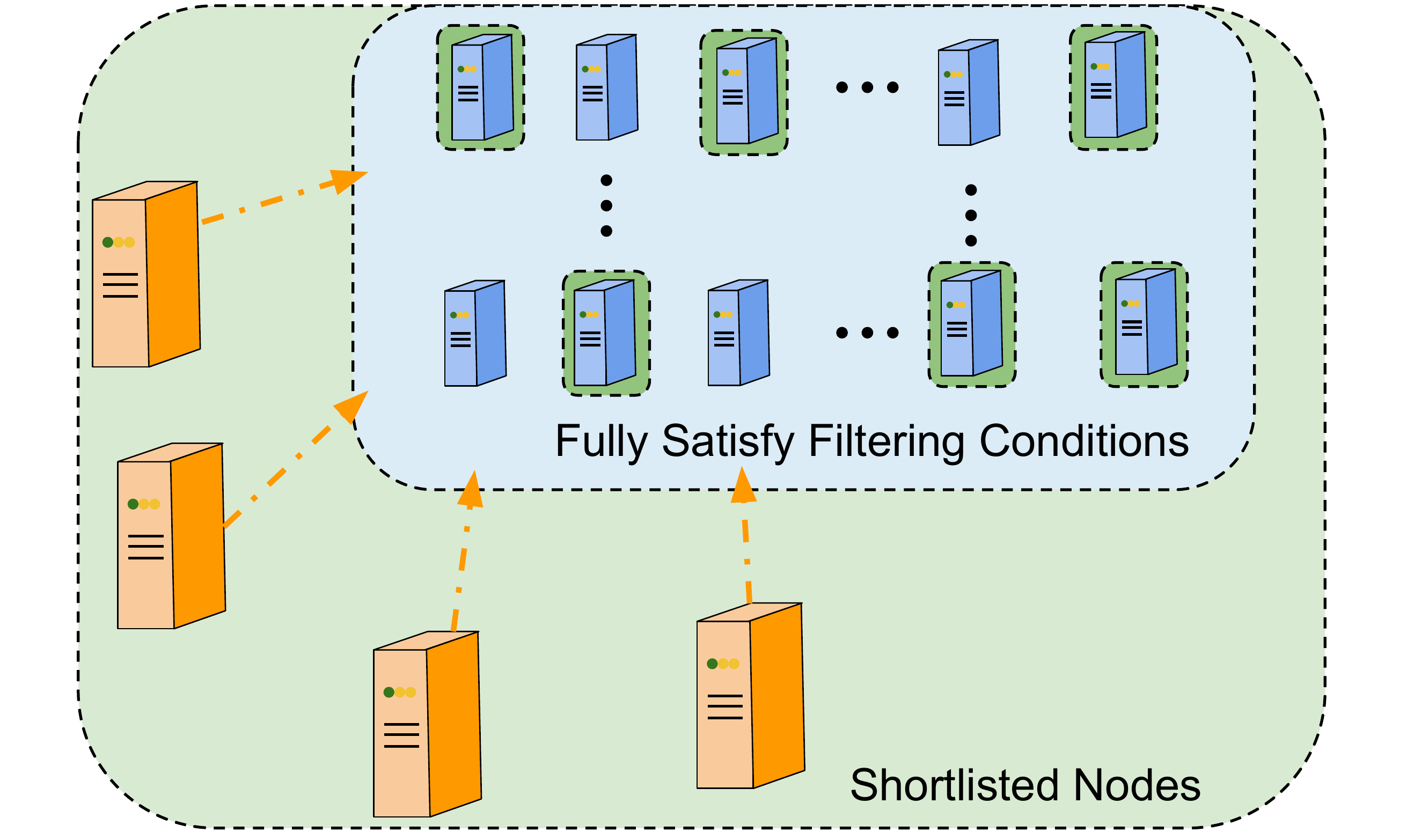}
\caption{The diagram of our mitigation strategy.}\label{FigMitigationMethod}
\end{figure}
The affinity features make schedulers' outcomes more predictable and can hence be manipulated by malicious attackers. To defend against {\proj}, a straightforward method is to re-introduce randomness into the scheduling process. To this end, we manually insert nodes that do not pass the filtering phase of the scheduling process to the scoring candidate lists. This comes with violations of user requirements, but according to our later results, the violation is not severe.

The proposed mitigation method is to randomly skip label check in the filtering step of scheduling with a specified probability $p_s (0 \leq p_s \leq 1)$. This can be considered loosening the filtering conditions and manually increasing the number of available nodes to schedule user applications. This intentional violation of user specifications comes with the risk of violating user-specified requirements and can possibly cause performance degradation. Still, according to our results in this section the sacrifice is relatively acceptable compared to the benefits it brings in defending against {\proj}. We will present in later that with only a relatively small sacrifice of violating user requirements, the co-location rate of attacks is significantly reduced. The performance degradation can be interpreted as the cost for security. 

The detailed pseudo-code of our mitigation method is shown in Algorithm \ref{AlgMitigation}. We introduce a simple \textsc{Skip} function to help determine whether or not a label checking step should be omitted, and use \textsc{Skip} function to modify the \textsc{Filter} function in Algorithm \ref{AlgScheduling}. As can be seen from Algorithm \ref{AlgMitigation}, this mitigation method is relatively easy to integrate into existing systems, and there is a parameter ($p_s$) that can be adjusted to control the level of defense in the system.
\begin{algorithm}[htbp]
\caption{Mitigation strategy.}\label{AlgMitigation}
\begin{algorithmic}
\Function{Skip}{$p_s$}
	\State Generate a $1$ bit $flag$. $flag$ is $1$ with probability $p_s$.
	\If{$flag = 1$}
		\State \Return True
	\Else 
		\State \Return False
	\EndIf
\EndFunction

\Function{NewFilter}{$userspecs$, $allnodes$, $p_s$}
\State $candidates$ = $allnodes$
\For{$node$ in $allnodes$}
	\For{$spec$ in $userspecs$}
		\If{$spec$ is a resource specification \\
		\hspace{1.5cm} or \textsc{Skip($p_s$)} == False}
			\If{$spec$ is not satisfied on $node$}
				\State Delete $node$ from $candidates$
			\EndIf
		\EndIf
	\EndFor
\EndFor
\State \Return $candidates$
\EndFunction
\end{algorithmic}
\end{algorithm}

It is worth noting that scheduler designers should be careful about skipping user-specified label checks since simply ignoring some of them will cause the application to be scheduled to nodes without necessary hardware or system settings. The choices of labels to skip ought to be limited to those only affecting performance to guarantee correct execution. One example is if the application requires a special type of hardware architecture, this specification cannot be ignored.

\subsection{Effects of Mitigation Strategy}
\begin{figure*}[htbp]
\centering
     \begin{subfigure}[t]{0.45\linewidth}
         \centering
         \includegraphics[width=\textwidth]{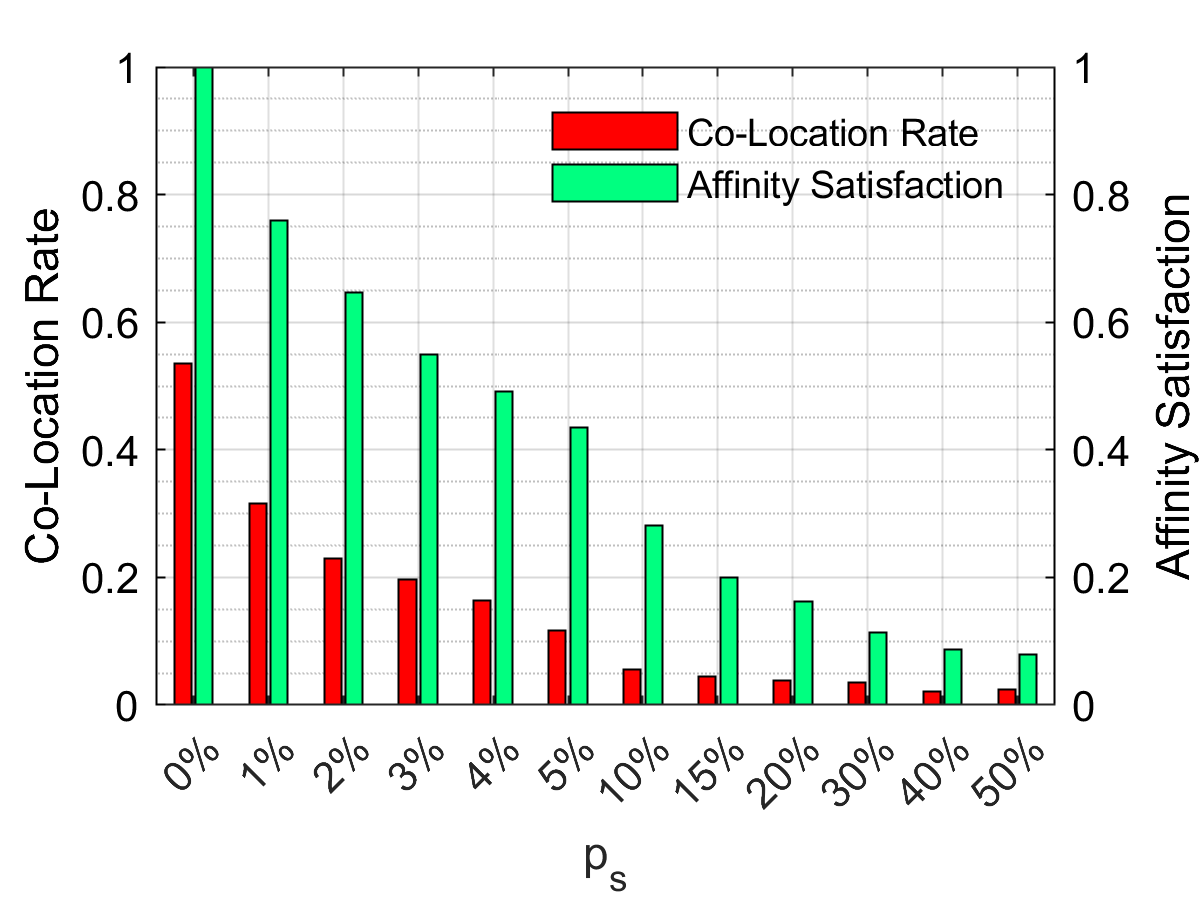}
         \caption{$p_{mn}=0.5$, $p_{ma}=0.5$.}
         \label{FigMitigation:a}
     \end{subfigure}
     \hfill
     \begin{subfigure}[t]{0.45\linewidth}
         \centering
         \includegraphics[width=\textwidth]{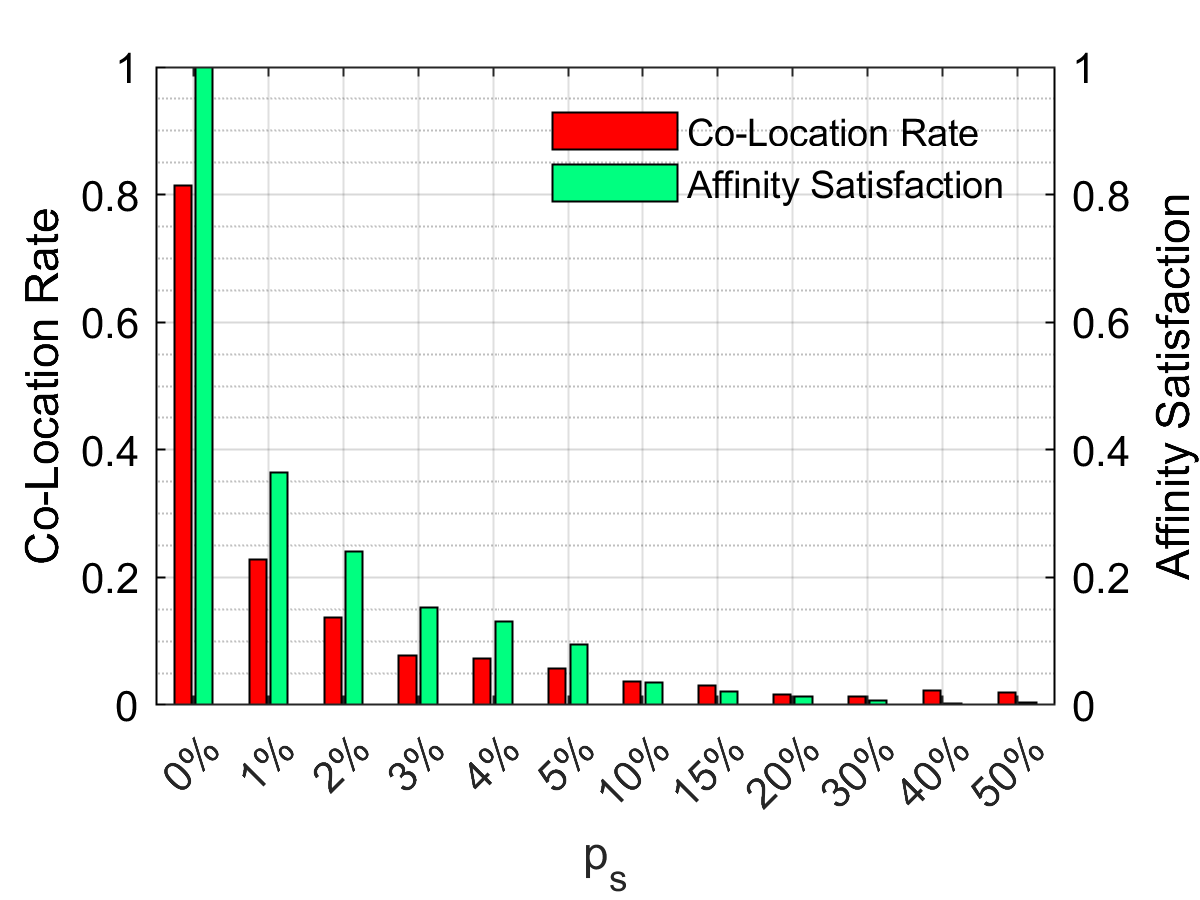}
         \caption{$p_{mn}=0.9$, $p_{ma}=0.9$.}
         \label{FigMitigation:b}
     \end{subfigure}
     \hfill
      \begin{subfigure}[t]{0.45\linewidth}
         \centering
         \includegraphics[width=\textwidth]{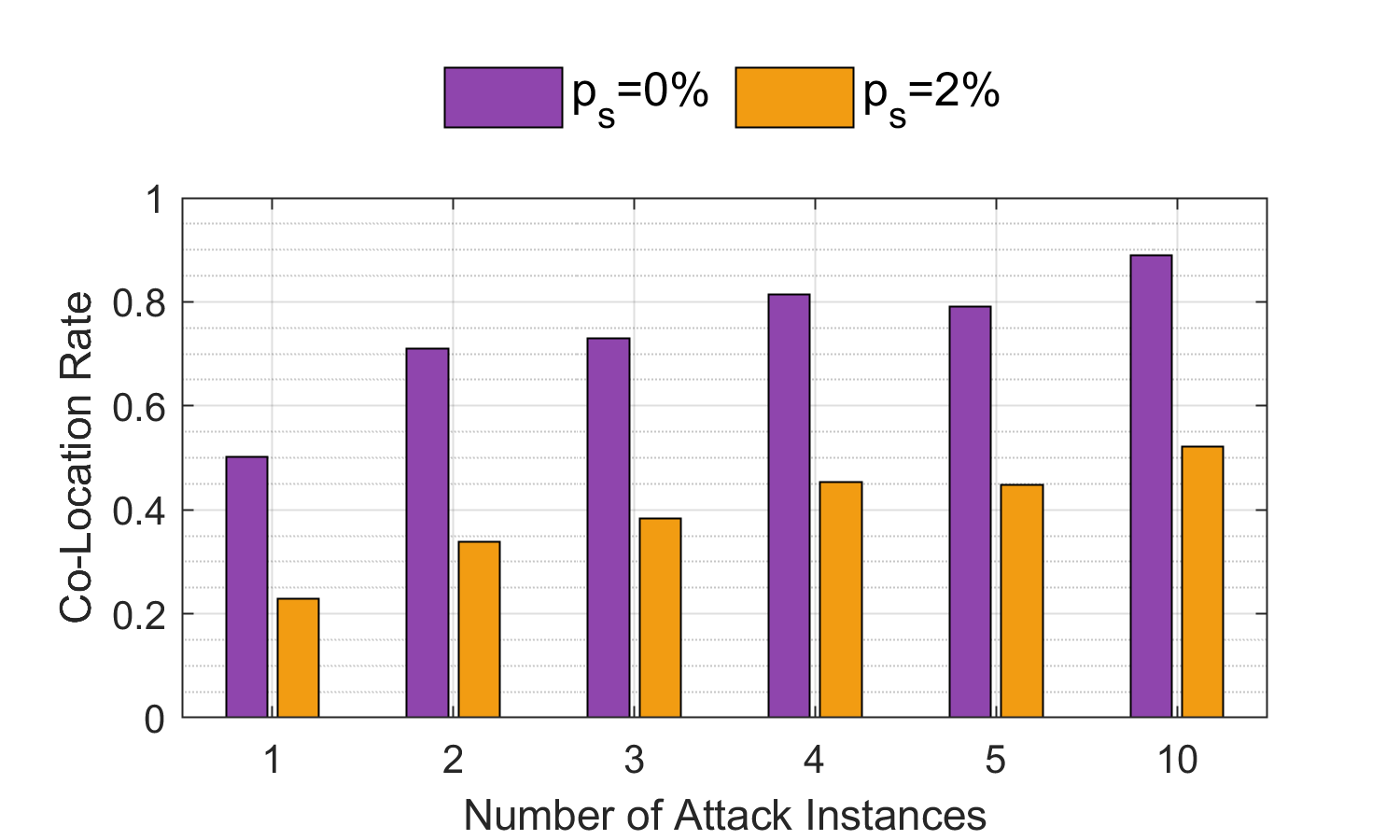}
         \caption{Co-location rates before and after our mitigation strategy is applied to attacks with different number of attack instances. $p_s$ is set to $2\%$.}
         \label{FigMitigation:c}
     \end{subfigure}
     \hfill
     \begin{subfigure}[t]{0.45\linewidth}
         \centering
         \includegraphics[width=\textwidth]{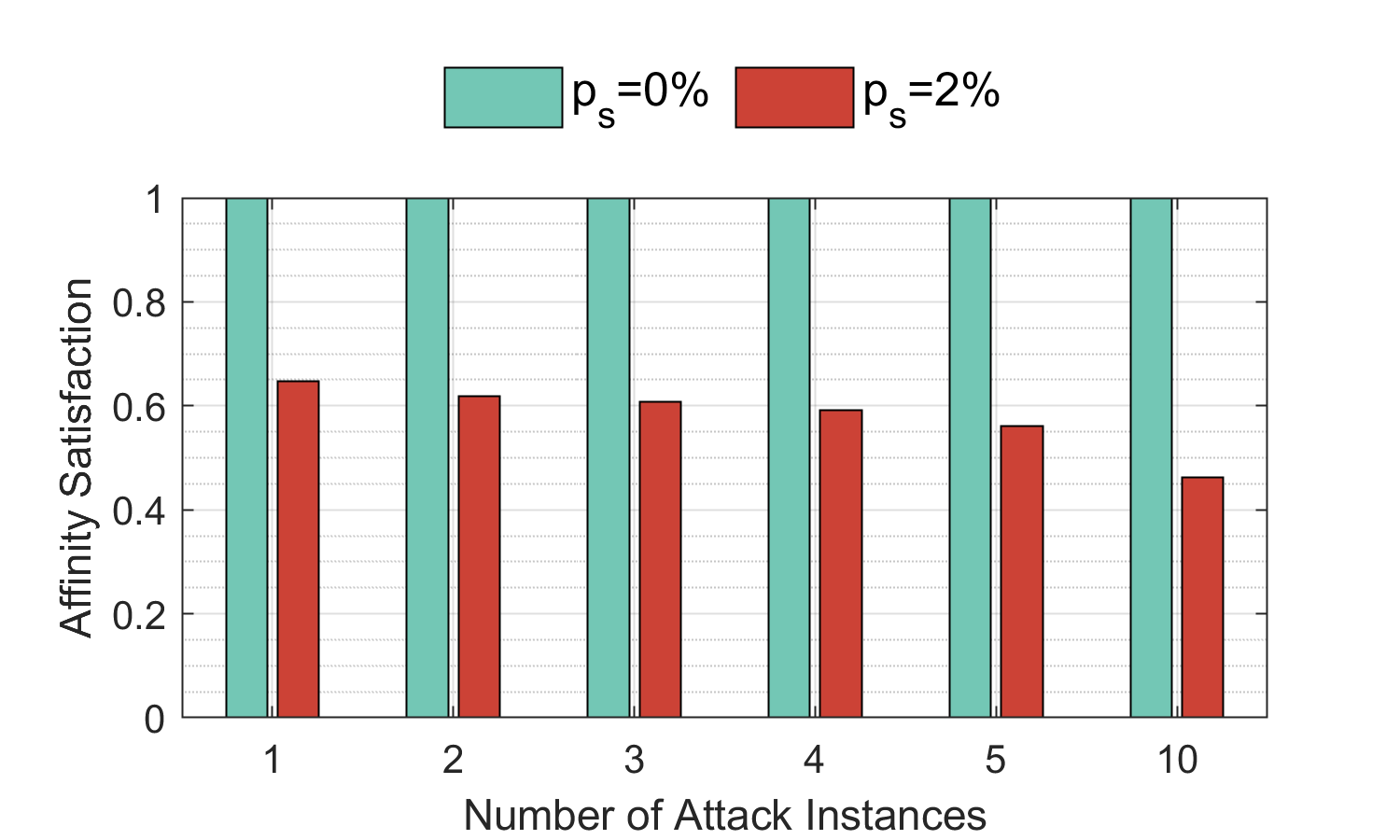}
         \caption{Affinity satisfaction scores in the experiments.}
         \label{FigMitigation:d}
     \end{subfigure}
     \caption{Co-location results after introducing migration.}
        \label{FigMitigation}
\end{figure*}

We test our mitigation strategy in simulation and present the results in Figure \ref{FigMitigation}. Since it is hard to model the actual performance as it varies according to cluster settings and it is highly related to the heterogeneity of the target cluster, we use a metric called affinity satisfaction to quantify our sacrifice in performance. Since a lot of these affinity specifications are provided by users for the sake of performance, and our strategy is to ignore part of these specifications randomly, this metric can reflect the trade-off between performance and security. After making a scheduling decision, we re-run the filtering check on the node the task is assigned to and record whether all affinity requirements are met. After the entire experiment finishes, we calculate the percentage of applications assigned to a node that fully satisfies all their required affinity and use that as our metric.

Figure \ref{FigMitigation:a} and Figure \ref{FigMitigation:b} show how our mitigation strategy reduces the co-location risk and the sacrifice it brings. We vary $p_s$, the probability of skipping affinity check of a label and re-run the simulation in Section \ref{SubSecNumAttackSim}. In Figure \ref{FigMitigation:a}, with $p_s=5\%$ the co-location rate drops from $50\%$ to around $10\%$, while still keeping approximately $55\%$ of scheduling results that fully satisfy user needs. Please note that our metric only measures whether all requirements that aim to enhance performance are fully satisfied. As shown in Table \ref{TabViolatedAffinity}, in a lot of the cases, there may be only a limited number of affinity specifications violated; hence the applications will still be able to have most of their requirements met. In Figure \ref{FigMitigation:b} we simulate an environment that users use affinity features more extensively. Comparing to Figure \ref{FigMitigation:a}, we conclude that when users utilize more such features, the co-location rate and the affinity satisfaction metric both become more sensitive to our mitigation strategy. This is easy to understand as well. The scheduler outputs rely more on user specifications as users use those features more extensively, but our strategy randomly forces the scheduler to give up user specifications in the filtering step. Therefore more user specifications will be violated as $p_s$ increases, resulting in a quicker decrease in co-location rate as well as affinity satisfaction metric.

In Figure \ref{FigMitigation:c} and Figure \ref{FigMitigation:d}, we compare co-location rates before and after involving our mitigation method on attacks with different numbers of attack instances and provide the affinity satisfaction scores, respectively. Here, all attack instances are submitted with the spreading label to increase the co-location rate. As shown in Figure \ref{FigMitigation:c}, with $p_s=2\%$, our mitigation strategy reduces the co-location rate by at least around $15\%$. However, as the number of attack instances increases, the reduction becomes less significant. According to the result of Figure \ref{FigMitigation:d}, the affinity satisfaction metric does not change much as the number of attack instances increases. This proves that the metric of affinity satisfaction is only the function of $p_s$. With the results in Figure \ref{FigMitigation}, we conclude that bringing randomness to the scheduling process helps resist co-location attacks. But to defend against malicious attackers who issue multiple attack instances at a time, we will need to raise up $p_s$ to achieve the same reduction in co-location rate.

\begin{table*}[!t]
\renewcommand{\arraystretch}{1.3}
% if using array.sty, it might be a good idea to tweak the value of
% \extrarowheight as needed to properly center the text within the cells
\caption{The average number of violated affinity specifications. In our experiments, since $10$ effective labels are used (excluding the one used as spreading label for attackers), the average number of affinity specifications for $p_{m_n}=p_{m_a}=0.5$ is $5$ and the average number of affinity specifications for $p_{m_n}=p_{m_a}=0.9$ is $9$.}
\label{TabViolatedAffinity}
\centering
% Some packages, such as MDW tools, offer better commands for making tables
% than the plain LaTeX2e tabular which is used here.
\begin{tabular}{|c||c|c|c|c|c|c|c|c|c|}
\hline
$p_{m_n}$, $p_{m_a}$ & $p_s=0\%$ & $p_s=1\%$ & $p_s=2\%$ & $p_s=3\%$ & $p_s=4\%$ & $p_s=5\%$ & $p_s=10\%$ & $p_s=15\%$ & $p_s=20\%$\\
\hline
$0.5$ & $0.00$ & $0.45$ & $0.68$ & $0.88$ & $1.07$ & $1.19$ & $1.68$ & $2.00$ & $2.17$\\
\hline
$0.9$ & $0.00$ & $1.65$ & $2.29$ & $2.78$ & $3.02$ & $3.33$ & $4.04$ & $4.40$ & $4.57$\\
\hline
\end{tabular}
\end{table*}

According to the experimental results, the sacrifice on affinity matching is not significant. Also, since all basic resource requirements are guaranteed to meet because we don’t change the CPU and memory allocation strategy, and our mitigation strategy only violates a relatively small number of affinity requirements, we believe the performance degradation is not much.

\section{Discussion}\label{SecDiscussion}
According to our experimental results in Section \ref{SecExperiments}, security problems mainly arise from user specifications that aim to fine-tune the scheduling decisions by specifying preferences or requirements on nodes and accompanying applications. In heterogeneous clusters, giving users access to control the output of the schedulers is possible to expose such security flaws to co-location attacks. These specifications help narrow down the search space of available nodes in the filtering step and force the scheduler to score the target machine with higher scores. In previous research works, co-location is achieved via issuing a large number of attacks, which is relatively easy to detect and prevent. When enabling user specifications in scheduling, by only issuing a few attack instances, the co-location rate is relatively high. From the results shown in Section \ref{SecExperiments}, we have summarized guidelines for attackers as well as regular users and cluster managers as detailed below.

\begin{figure}
\centering
\includegraphics[width=\linewidth]{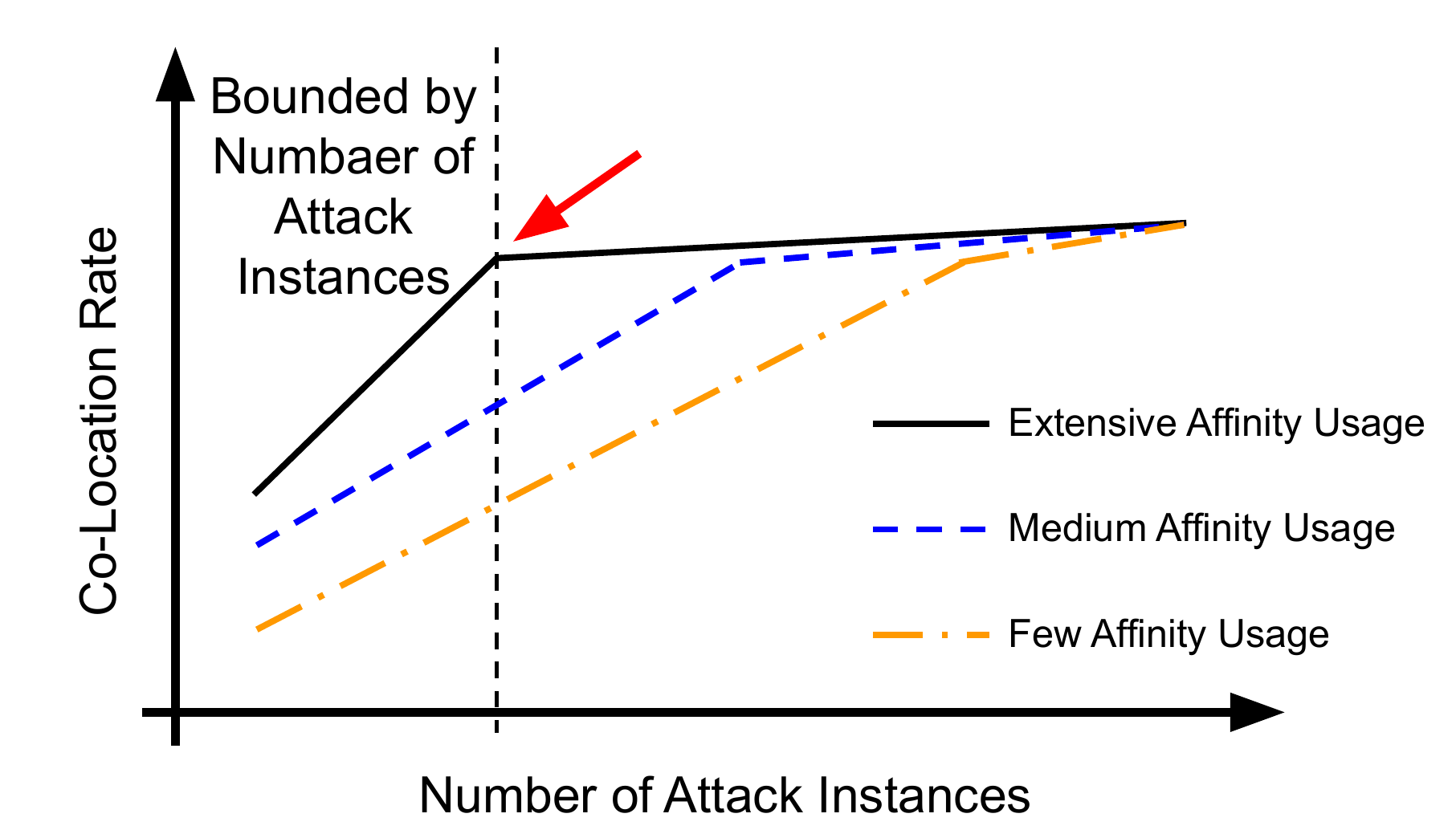}
\caption{The roofline model of the influence of number of attack instances.}\label{FigRoofline}
\end{figure}

For attackers, the first thing they should do is to study the target applications and have a guess on the type of requirements the victim user will specify. To increase coverage, a malicious user can issue multiple versions of their attack applications, which cover different possible settings that users might provide. As for the number of attack instances they should launch, according to our experimental results in Section \ref{SecExperiments} there is a best trade-off point between co-location accuracy and cost (as shown in Figure \ref{FigRoofline}). Before reaching this point, increasing the number of attack instances will raise the co-location rate. In contrast, beyond that point, issuing more attack instances (more cost) for a single target brings a relatively low gain in co-location rate. 

For users, it is highly recommended that they use as few affinity features as possible to avoid predictable, mimicable scheduling results. Also, their specifications on their applications should be kept secret and avoid being leaked to malicious attackers. The co-location rate would be very low if few such specifications are copied by attackers.

For cluster managers, the key to defend against co-location attacks issued by malicious users is to bring randomness into the scheduling process and provide users with limited control over the scheduling results. Randomness may violate service level agreement (SLA); hence, it is crucial for cloud managers to make the trade-off between security and user specification satisfaction.
Also, when deploying their clusters and labeling all servers in the cluster, heterogeneity should be exposed to users as little as possible, making it harder for users to identify the nodes their applications are running on.

In some recent studies~\cite{delimitrou2014quasar,alipourfard2017cherrypick,venkataraman2016ernest}, instead of directly giving users the ability to control schedulers' outputs, the schedulers try to profile the applications users submit and use machine learning techniques to learn and find the best resource configuration and determine the best node for the application to run. The use of such adaptive techniques enhances the quality of scheduling, as the deployment of user applications is specially optimized. This, however, may also trigger some security problems since there are also methods to generate fake traces and trick the scheduler. By simply mimicking the behaviors of victim applications, the malicious attackers are also able to trick the scheduler of a heterogeneous cluster and get co-located with the victim application, according to some simulation results. Similar to our study, this problem occurs when schedulers try to find the most suitable machine according to the execution patterns, which inevitably narrows down the search space and greatly increases the chance of co-location.

The conflict between performance and the danger of co-location attack reveals a trade-off in scheduler design. According to the results in this paper, to provide security, randomness is a necessity; otherwise, once attackers obtain enough information about victims, co-location can be achieved with a high success rate. Randomness can be introduced explicitly (e.g., our method in Section \ref{SecMitigation}) or inexplicitly (e.g., by limiting the scheduler to consider only the current capacity of machines in the cluster because this information is not predictable by users). Our next step is to construct and test a scheduler with learning capabilities in a real large cluster and try to develop some mechanisms to provide secure scheduling results that also guarantee the performance of user applications.

\section{Related Work}\label{SecRelatedWk}
\subsection{Affinity Policies}
Affinity policies have been utilized in cloud systems to constrain instance placement and improve the scheduling quality. In 2012, a system called VMware’s Distributed Resource Scheduler (DRS)~\cite{gulati2012vmware} was proposed. DRS utilizes affinity policies to constrain placement of VMs and provide appropriate recommendations of nodes to users. The affinity policies in~\cite{gulati2012vmware} are called VM-to-Host rules and VM-to-VM rules, which correspond to the node affinity rules and inter-application affinity rules in this paper. In 2018, Moreno-Vozmediano \textit{et.al}~\cite{moreno2018orchestrating} implement a cloud orchestration system with new affinity policies like role-to-role rules and VM-to-location rules to address challenges in deploying high availability services. According to the analysis in this paper {\proj} can also exploit such features. Affinity rules widely exist in nowadays popular cloud architectures like Kubernetes~\cite{kubernetes}, OpenStack~\cite{openstack} and OpenNebula~\cite{opennebula}.

\subsection{Co-Location Attack}
The concept of co-location attack in the cloud was proposed and discussed in~\cite{ristenpart2009hey}. The authors analyze the risks of side-channel attacks by being co-located with malicious VMs in the public cloud. The paper also introduces the technique of network probing to determine whether the attack instances are co-located with victim instances, and the method the authors use to achieve co-location is issuing multiple attack VMs by brute force or exploit the scheduling locality regarding timing. Following this research, there are various works on the side-channel attack part~\cite{inci2016cache,zhang2014cross, cardenas2012detection} as well as the co-location detection part~\cite{zhang2011homealone,inci2016co,sullivan2018microarchitectural,inci2015seriously}. Regarding the co-location step, researchers normally utilize brute-force issue of attack instances~\cite{han2015using,ristenpart2009hey} or exploit the scheduling locality in time to spread attack instances in the cloud. The paper~\cite{han2015using} formalizes this question and provides the expression of minimum VMs the attacker needs to issue under different scheduling patterns. The results shown in the paper~\cite{varadarajan2015placement} is relevant to ours, where the authors examine the effects of different factors on co-location rates on commercial cloud providers.

Researchers have studied mitigation methods to prevent co-location.~\cite{azar2014co} provides a simple algorithm that brings randomness to the scheduling process and mathematically proves the reduction in security risk.~\cite{han2020quantify} employs a machine-learning-based method to identify malicious user patterns.~\cite{zhang2012incentive} presents the usage of migration of VMs to reduce co-location risk.~\cite{dhavlle2021imitating} proposed to add perturbations that prevent co-location detection attempts. 

A concurrent work proposed Cloak \& Co-locate attack~\cite{makrani2021cloak}, which targets profiling-based schedulers to achieve co-location by providing misleading execution traces. However, to the best of our knowledge, such schedulers were only studied in the research community~\cite{delimitrou2014quasar,delimitrou2013paragon} and is not currently deployed by the industry. Furthermore, our work targets the widely deployed filter-score based schedulers and achieves co-location by replicating user specified constraints.

To the best of our knowledge, we are the first to explore the security risks of allowing users to submit scheduling specifications along with the instances to the cloud. Also, when these features are utilized, the mitigation methods may not be as effective since the scheduler still obeys the rules the users provide.

\section{Conclusion}\label{SecConclusion}
In this paper, we present that the filter-score scheduler is prone to suffer from security problems because of the use of the existing affinity features. We find that, if the affinity features are utilized by users in the system, by some simple guesses according to the application information or obtaining the submitted configuration files, the attacker can relatively accurately locate the victims to achieve co-location and proceed to issue side-channel attacks. We propose a simple attack strategy called {\proj} and a corresponding mitigation method to defend against it. The effectiveness of {\proj}, as well as the proposed mitigation method, are supported by our experimental results in either the CloudLab cluster or simulation. The findings in this paper provide several guidelines regarding security for attackers, users, and, most importantly, cluster managers or designers. From the perspective of cluster scheduler designers, for safety concerns, we advise that the cluster information should be kept secret, and scheduling decisions should be made with user interference at the least level, where there is a trade-off between security and performance. Profiling applications and utilize machine learning techniques to determine the best configuration for an application can be one method to hide details of clusters yet still keep performance, but risks still exist since the behaviors of an application can be mimicked, and the scheduler can be tricked. Future work will be dedicated to how micro-architectural attacks can be initiated on the cloud and how to design schedulers that can reach a well-balanced point.

%One interesting research question here is how to determine whether the attack instances are successfully co-located with the victim, thus the following operations (shutdown/keep running) can proceed. Another research question here is how to disguise side-channel attack instances and bypass the defense methods in cloud. This part is not the focus of this paper and we will present related results in future work.
\section*{Acknowledgment}
The authors would like to thank the support of High Performance Computing (HPC) Center of University of California, Davis, and Mr. Sergey Buduchin, the technical director of HPC Center. We are also grateful to our shepherd Mauro Conti and all anonymous reviewers for their constructive feedback.



\bibliographystyle{IEEEtranS}
\bibliography{main.bib}


% \end{thebibliography}

\end{document}